\documentclass[a4paper,fleqn,usenatbib]{mnras}

\usepackage{newtxtext,newtxmath}

\usepackage[T1]{fontenc}
\usepackage{ae,aecompl}

\usepackage{graphicx}	
\usepackage{amsmath}	
\usepackage{amssymb}	
\usepackage{slashed, epsf, color}
\usepackage{mathrsfs}
\usepackage{url}
\usepackage{comment}
\usepackage{ifthen}
\usepackage{xspace}
\usepackage{epstopdf}
\newcommand{\unit}[1]{\,\text{#1}}
\newcommand{\Order}[1]{{\cal O}(#1)}

\newcommand{\mDM}{m_{\text{DM}}}
\newcommand{\rhoDM}{\rho_{\text{DM}}}

\newcommand{\red}[1]{{#1}}

\newcommand{\errs}[4]{${#1}_{-#2}^{+#3}\pm{#4}$}

\newcommand{\UI}{Ursa\,Major\,I\xspace}
\newcommand{\UII}{Ursa\,Major\,II\xspace}
\newcommand{\Seg}{{Segue}\,1\xspace}
\newcommand{\Coma}{Coma\,{Berenices}\xspace}
\newcommand{\Ls}{{\cal L}_\text{s}}
\newcommand{\Lm}{{\cal L}_\text{m}}
\newcommand{\imax}{i_{\text{max}}}

\newcommand{\thetaROI}{\theta_{\text{ROI}}}
\newcommand{\vFGj}{v_{\text{FG}j}}
\newcommand{\vFGjo}{v_{\text{FG0}j}}

\newcommand{\sigmaFGj}{\sigma_{\text{FG}j}}
\newcommand{\sigmaFGjo}{\sigma_{\text{FG0}j}}

\newcommand{\KI}{KI17\xspace}
\newcommand{\Cont}{{\sl Contaminated}\xspace}
\newcommand{\MS}{{\sl Conventional}\xspace}
\newcommand{\JIn}{J_{\text{Input}}}

\title[Foreground effect on the $J$-factor estimation]{
Foreground effect on the $J$-factor estimation\\
 of ultra-faint dwarf spheroidal galaxies
}

\author[K. Ichikawa et al.]{
Koji Ichikawa,$^{(a)}$
Shun-ichi Horigome$^{(a)}$
Miho N. Ishigaki,$^{(a)}$
Shigeki Matsumoto,$^{(a)}$\newauthor
Masahiro Ibe,$^{(a, b)}$
\ Hajime Sugai,$^{(a)}$
and 
Kohei Hayashi$^{(b)}$
\\
$^{(a)}${Kavli IPMU (WPI), UTIAS, The University of Tokyo, Kashiwa, 277-8583, Japan} \\
$^{(b)}${ICRR, The University of Tokyo, Kashiwa, 277-8582, Japan} \\
}

\date{Accepted XXX. Received YYY; in original form ZZZ}

\pubyear{2017}

\begin{document}
\label{firstpage}
\pagerange{\pageref{firstpage}--\pageref{lastpage}}
\maketitle

\begin{abstract}
Dwarf spheroidal galaxies (dSphs) are promising targets for the gamma-ray dark matter (DM) search. 
In particular, DM annihilation signal is expected to be strong in some of the recently discovered nearby ultra-faint dSphs, 
which potentially give stringent constraints on the $\Order{1}$ TeV WIMP DM. 
However, {various} non-negligible systematic uncertainties complicate the estimation of the astrophysical factors relevant for the DM search in these objects. 
Among them, the effects of foreground stars particularly attract attention 
because the contamination is unavoidable even for the future kinematical survey. 
In this article, we assess the effects of the foreground contamination on the astrophysical $J$-factor estimation 
by generating mock samples of stars in the four ultra-faint dSphs and using a model of future spectrographs. 
We investigate various data cuts to optimize the quality of the data and apply a likelihood analysis which takes member and foreground stellar distributions into account. 
We show that the foreground star contaminations in the signal region (the region of interest) 
{and their statistical uncertainty can be estimated}
by interpolating the foreground star distribution in the control region 
where the foreground stars dominate the member stars. 
Such regions can be secured at future spectroscopic observations
utilizing a multiple object spectrograph with a large field of view; 
e.g. the Prime Focus Spectrograph mounted on Subaru Telescope. 
The above estimation has several advantages: 
The data-driven estimation of the contamination makes the analysis of the astrophysical factor stable against the complicated foreground distribution. 
{Besides, foreground contamination effect is considered in the likelihood analysis.}
\end{abstract}




\section{Introduction}
\label{sec:intro}

Various astrophysical observations such as 
the dynamics of galaxy clusters\,\citep{1933AcHPh...6..110Z}, 
rotation curves of spiral galaxies\,\citep*{1978ApJ...225L.107R,1980ApJ...238..471R}, and
gravitational lensing\,\citep{McLaughlin:1998sb, Lokas:2003ks,Clowe:2006eq,Bradac:2006er},
strongly indicate the existence of  dark matter (DM) in the astronomical objects.
{A} recent global fit of 
the Cosmic Microwave Background (CMB), 
Large Scale Structure (LSS), and 
Supernovae (SNe) observations\,\citep{Ade:2015xua}
reveal that quarter of the total energy of the universe consists of DM.
One of the most attractive candidates of DM is weakly interacting massive particle (WIMP), 
which naturally explains the observed dark matter density with its annihilation channels into lighter standard model particles. 
Particularly, the WIMP dark matter with $\lesssim \Order{1}$\,TeV 
has drawn attention  in the context of the physics beyond the standard model such as supersymmetry (see e.g. \citealp*{Jungman:1995df} also \citealp{Murayama:2007ek,Feng:2010gw}).

Gamma-ray indirect detection experiment, which aims to observe gamma-rays induced by the DM annihilation, has a strong sensitivity to this $\Order{1} \unit{TeV}$ WIMP.
Among various astronomical objects, dwarf spheroidal satellite galaxies (dSphs) associated with the Milky Way are the ideal targets due to its small distance ($\sim 10\--{\text{a few hundred}} \unit{kpc}$ from the solar system) and dense DM environment with low astrophysical background\,{\citep{PhysRevD.86.023528, Lefranc2016}}.
However, recent studies
show that expected signal flux coming from the dSphs is significantly affected by various uncertainties such as 
the statistical procedure\,\citep{Martinez:2009jh}, 
DM distribution \,\citep{Geringer-Sameth:2014yza, Bonnivard:2015xpq, Hayashi:2016kcy}, 
stellar distribution\,\citep{Ullio:2016kvy}, 
unresolved binary stars\,(
\citealp*{Mateo:2007xh}; \citealp{Koch:2007ye, Minor:2013bj, Simon:2007dq,Simon:2010ek,McConnachie:2010pn,Koposov:2011zi,Kirby:2013isa,Simon:2015fdw})
and foreground contamination \citep{Bonnivard:2015vua, 2017MNRAS.468.2884I}.

Although future deep spectroscopic surveys would mitigate most of these systematic uncertainties,
the foreground contamination remains problematic
because the fraction of the foreground would not be suppressed or even become worse in the future observation.
{
In \citet{2017MNRAS.468.2884I} (hereafter \KI),
we have investigated the effects of the foreground contamination for classical dSphs. 
We found that, when simple color-magnitude, velocity, and surface-gravity cuts are employed to select the member stars, 
in which case the foreground fraction is less than 5\%, 
the contamination can lead to an overestimation of the signal flux by a factor of $\sim 3$.
}
In \KI, we resolve this foreground effect by introducing a multi-component fit in which the distribution{s} of the 
member stars and foregrounds are mixed.

The situation is more problematic for ultra-faint dSphs (UFDs).
The {UFD}s {were} discovered after SDSS\,II and contains smaller number of the stars inside the system.
Although recent kinematical analyses\,\citep{Bonnivard:2015xpq,Ackermann:2015zua,Geringer-Sameth:2014yza} indicate 
that the signal flux coming from the UFDs can be much stronger than that from the classical dSphs,
the uncertainties of these signal fluxes are much larger 
due to the lack of the knowledge of the kinematics inside the system.
In particular, a recent study\,\citep{Bonnivard:2015vua} reveals 
that the foreground contamination can significantly affect the estimation of the signal flux 
by two orders of magnitude at most.
Therefore, precise analysis of the foreground effect for the UFDs is required 
and will play an essential role in the future deeper spectroscopic surveys.

In this paper, we test the foreground effect for the UFDs 
by generating realistic stellar mock data and applying the likelihood analysis developed in \KI.
We also compare the results with those obtained by the other conventional analyses.
The organization of this paper is as follows.
In Sec.\,\ref{sec:flux}, we review the formula of the gamma-ray signal flux
{and defines the so-called $J$-factor}.
In Sec.\,\ref{sec:analysis}, we provide the procedure of our analysis.
The results of the fits are given in Sec.\,\ref{sec:results}.
Finally, we summarize our discussion in Sec\,\ref{sec:summary}.

\section{Signal Flux and $J$-factor}
\label{sec:flux}

The gamma-ray signal flux of DM annihilation {stemmed} from 
the dSphs can be expressed by the following formula:
\begin{eqnarray}
\Phi (E, \Delta \Omega)
=
\left[
\frac{C \langle \sigma v \rangle}{4 \pi \mDM^2}
\sum_{f}
b_{f}
\left(\frac{dN_\gamma}{dE}\right)_f
\right]
\times J(\Delta \Omega)
\ .
\label{eq:fluxformula}
\end{eqnarray}
The coefficient $C$ is 1/2 for Majorana and 1/4 for Dirac dark matter.
Dark matter mass is defined by $\mDM$.
The product of the total annihilation cross section $\sigma$ and the relative velocity $v$ 
is averaged with the velocity distribution function (represented by $\langle \dots \rangle$). 
The branching fraction of the annihilation channel $f$ is denoted by $b_f$, 
while the differential number density of photons from a given final-state $f$ is given by 
$(dN_\gamma/dE)_f$.

The factor $J$ after the parenthesis in the right-hand side (so-called $J$-factor)
reflects the amount of the squared DM density 
inside the cone with a solid angle $\Delta \Omega$:
\begin{eqnarray}
J (\Delta \Omega) = 
\left[
\rule{0ex}{4ex}
\int_{\Delta \Omega} d\Omega \int_{l.o.s.} dl \,
\rho^2(l, \Omega)
\right]\ .
\end{eqnarray}
Here we define the dark matter profile at a distance $l$ and angle $\Omega$ by $\rho(l, \Omega)$.
The integration of $l$ is performed along the line-of-sight.

{
Currently, the most common way of the DM profile estimation is to apply dynamical mass models based on Jeans equations to the line-of-sight velocity of stars in the dSph.
Under the assumption of the spherical symmetry for luminous and dark components of the dSph, the projected velocity dispersion at a projected radius $R$ can be written by
}
\begin{eqnarray}
\sigma^{2}_{l.o.s}(R) = \frac{2}{\Sigma_{*}(R)} 
\int^{\infty}_{R}  dr
\left( 1 - \beta_{\text{ani}} (r) \frac{R^2}{r^2} \right)
\frac{\nu_{*} (r) \sigma^{2}_{r} (r)}{\sqrt{1 - R^2/r^2}}
\ ,
\label{eq:displos}
\end{eqnarray}
where 
$r$ denotes the {un-projected} distance from the centre of the dSph,
 and $\Sigma_{*}(R)$ is the projected spatial stellar distribution obtained by integrating the stellar distribution $\nu_{*} (r) $ along the 
projected direction.
The anisotropy parameter $\beta_{\text{ani}}$ is defined by
$\beta_{\text{ani}} = 1 - (\sigma^{2}_{\theta}/\sigma^{2}_{r})$
where we define the radial, azimuthal, and  polar components of the 3-dimensional dispersion curve 
as $\sigma_{r}$, $\sigma_{\theta}$, and $\sigma_{\phi}$, {respectively, in a spherical coordinate} and take $\sigma_{\theta}  =  \sigma_{\phi}$
for the spherical symmetry.

{
The radial dispersion curve, $\sigma^2_r$, is related to the gravitational potential (i.e. dark matter profile) 
through the spherical Jeans equation\,{\citep{2008gady.book.....B}}.
}
Under the assumption of constant $\beta_{\text{ani}}$,
{this dispersion curve}
can be expressed as\,\citep{1994MNRAS.270..271V,2005MNRAS.363..705M}
\begin{eqnarray}
\sigma^{2}_{r}(r) = 
\frac{1}{\nu_{*} (r) }
\int^{\infty}_{r} 
\nu_{*} (r')
\left(\frac{r'}{r}\right)^{2 \beta_{\text{ani}}}
\frac{G M(r')}{r'^2} 
dr'\ .
\label{eq:dispr}
\end{eqnarray}
Here $G$ is the gravitational constant, and $M(r)$ is the enclosed mass of the {spherical} dark matter halo{:} 
$M(r) \equiv \int^{r}_{0} 4 \pi r'^2 {\rhoDM} (r') dr'$. 
From Eq.\,(\ref{eq:displos}) and Eq.\,(\ref{eq:dispr}), 
we can estimate the DM profile {$\rhoDM$} by constructing the dispersion curve $\sigma_{l.o.s.} (R) $
{from the observational data.} 

As we have discussed in \KI, 
the dominant uncertainty of the signal flux comes from the $J$-factor.
This is because 
while the parenthesis in Eq.\,(\ref{eq:fluxformula}) is well controlled by the calculation of particle physics,
the estimation of the $J$-factor is limited by the number of the kinematical stellar data of the dSphs.
Although {the uncertainty} of the $J$-factor is still under discussion,\footnote{
This is due to the various biases in the estimation:
the statistical procedure, DM halo model,  
stellar distribution, 
unresolved binaries and foreground contamination,
as reviewed in \KI.
}
the error bar can be a few orders of magnitude larger for the UFDs, {which we focus on in this work}.
To suppress both the statistical and {systematic} uncertainties,
future deep spectroscopic observation is mandatory.

\section{Analysis}
\label{sec:analysis}

In this section, we introduce the mock-based analysis developed in \KI.
In our analysis, we first generate realistic mock dSph stellar data including foreground stars.
We sample this stellar data by accounting for a spectroscopic capability, 
which provides realistic mock samples of a future observation.
We next {attempt to} decrease the foreground fraction by imposing a selection rule. 
In this paper, we consider two approaches: naive cuts and selection by using the membership probability.
Finally, we perform the halo profile estimation by using two types of  the likelihood functions,
which have single and mixed component(s) in their distribution function respectively.
In Sec.\,\ref{sec:results}, we will provide the results of the analyses by three combinations of the selections and fits:
naive cut + mixed component fit, 
membership selection + single component fit,
and naive cut + single component fit.
{They} correspond to the \KI, {conventional} and the most naive approaches, respectively.

\subsection{Mock dSphs}
\label{subsec:mock}

\begin{table*}
\begin{center}
  \begin{tabular}{cccccccccc}
Model dSph
& $d$ [kpc]
& $r_e$ [pc]
& $\log_{10}\left(\frac{\rho_{s}}{[M_{\bigodot}/\text{pc}^3]}\right) $
& $\log_{10}\left(\frac{r_s}{[\text{pc}]}\right)$
& $\alpha$ 
& $\beta$ 
& $\gamma$ 
& $-\log_{10} (1-\beta_{\text{ani}})$ 
& $\log_{10} \left(\frac{\JIn}{[\text{GeV}^2/\text{cm}^5]}\right)$
\\ \hline
\UII & 32 & 149 & -0.370& 2.62& 2.36& 3.28& 0.0328& -0.975 & 19.70\\
\Coma & 44 & \red{77} &-0.283 & 2.27 & 2.87 & 6.79 & 0.178 & {-}0.894 & 18.74\\
\Seg & \red{23} & 29 & 0.306& 1.93& 0.973& 3.94& 1.15& -0.00155 & \red{20.05}\\
\UI  & 97 & \red{319} & 0.587& 1.97& 2.89& 8.04& 0.302& -0.625& \red{18.92} \\
\hline
 \end{tabular}
\caption{{\sl
The input parameters of each dSph.
The distances from the earth and projected half-light radii are shown by $d$ and $r_e$\,\citep{McConnachie:2012vd}.
The DM halo and kinematical parameters $\rho_s$, $r_s$, $\alpha$, $\beta$, $\gamma$, and $\beta_{\text{ani}}$ 
are determined by fitting the stellar data provided by
\citet{2007ApJ...670..313S} {for} \UII, \Coma, \UI and \citet{2011ApJ...733...46S} {for} \Seg under the same procedure as \citet{Geringer-Sameth:2014yza}.
\red{Quoted values are those obtained from the $\chi^2$ minimization.}\protect \footnotemark
The $\JIn$ shows the $J$-factors calculated within 
an angular radius of $0.5 \,\text{degree}$ under the input DM halo parameters and distance.
}}
\label{tb:dSphParameter}
\end{center}
\end{table*}
\footnotetext{\red{The slopes of the dSph DM profile are known to be poorly constrained,
so that the corresponding values in the table should be regarded as one of possible choices to generate merely mock data of UFDs.}}

As {models} of the mock dSphs,
we consider the four {UFD}s (\UII, \Coma, \Seg, and \UI),
in which the observation suggests abundant DM\,\citep{Hayashi:2016kcy,Bonnivard:2015xpq,Ackermann:2015zua,Geringer-Sameth:2014yza}.
\red{We use the same DM halo profile, the stellar distribution and the domain of parameters scanned as those of \citet{Geringer-Sameth:2014yza}}
based on the data provided by the kinematical observation{s} \citep{2007ApJ...670..313S,2011ApJ...733...46S}%
\footnote{{The data of \citet{2007ApJ...670..313S} was kindly provided by Josh Simon (private communication)}.}
and use the obtained DM profiles for the inputs of the dSph mocks.

In our analysis, the generalized dark matter halo density profile\,\citep{Hernquist:1990be,Dehnen:1993uh,Zhao:1995cp} is adopted as the input dark matter profile for the mock data and fit of the likelihood analysis:
\begin{eqnarray}
\rho_{\text{DM}}(r) = \rho_s (r/r_s)^{-\gamma} (1 + (r/r_s)^{\alpha})^{-(\beta-\gamma)/\alpha}\ ,
\label{eq:DMprofile}
\end{eqnarray}
where $r$ denotes the (un-projected) distance from the centre of the dSph{s},
 and parameters $\rho_s,\,r_s$ represent the typical density and scale of the halo respectively,
while parameters $\alpha,\,\beta,\,\gamma$ determine the shape of the halo density profile.
We also assume Plummer profile\,\citep{1911MNRAS..71..460P} for the member stellar distribution:
\begin{eqnarray}
\nu_{*}(r) = (3/4 \pi r_{e}^{3})\,(1 + (r/r_{e})^{2})^{-5/2} \ .
\label{eq:stellardist}
\end{eqnarray}
Here $r_{e}$ denotes the projected half-light radius of the dSph 
and we normalize the stellar distribution $\nu_{*}(r)$ to satisfy $\int 4 \pi r^2 \nu_{*}(r) dr = 1$.
The input parameters are shown in 
Table\,\ref{tb:dSphParameter}.\footnote{
In order to construct the kinematical data by using the method of \citet{1991MNRAS.253..414C} consistently,
we set the range of the anisotropy $\beta_{\text{ani}}$ to be $\beta_{\text{ani}} < 0$.
See, e.g., \citet{Ciotti:2009ic} for the limitation of the halo parameters
in the analytical solution of the Jeans equation.
}

The mock stellar data of each dSph is constructed  by assigning the colour, chemical abundance, and kinematical information.
Synthetic colour-magnitude diagrams are generated by utilizing the
PARSEC stellar isochrones\,\citep{2012MNRAS.427..127B} to represent 
observed properties of each dSph.
In detail, we first randomly draw a stellar initial mass from
the Salpeter initial mass function {(IMF)}. For that mock star, the age is
drawn from an uniform distribution in the
range 10$^{10.10}$-10$^{10.12}$ years, motivated by the fact that
the {UFD}s analysed in this work have been reported to be dominated by
an old stellar population\,\citep{2008AJ....135.1361D}. 
Similarly, the value of metallicity ([Fe/H]) is drawn 
from a Gaussian distribution with
{(mean, dispersion)$=$($-2.5$, $0.3$), ($-2.5$, $0.3$),
($-2.7$, $0.7$), and ($-2.2$, $0.6$) for
\UII, \Coma, \Seg, and \UI, respectively, which are}
approximately consistent with those estimated by \citet{2011ApJ...727...78K} and \citet{2010ApJ...723.1632N}.
Based on a theoretical isochrone for the given age and [Fe/H] values obtained
above, the absolute magnitude, colour and surface gravity
corresponding to the stellar initial mass are assigned.
The apparent magnitude and observed colour
are then calculated by adopting the distance
modulus from \citet{2012AJ....144....4M} and adding
typical photometric errors 
{increasing toward fainter magnitudes (0.012 at $i=20.0$ and 0.024 at $i=22.0$)}
as well as the Galactic extinction {from \citet{2011ApJ...737..103S}}.
At this point, the star is discarded if it is fainter
than the i-band limiting magnitude of 22.5. The mock stars are repeatedly
generated until the number of member stars brighter than the limiting
magnitude estimated by \citet{2008ApJ...684.1075M} is reached.
{
The stellar {IMF} can deviate from the Salpeter IMF below $M\sim0.5 M_\odot$ (see, e.g. the \citet{2001MNRAS.322..231K} IMF).
Magnitude limits ($i=21.0,\, 21.5$ and $22.0$) we adopted, however,
correspond to stellar masses well above 0.65 $M_{\odot}$,
and thus the mock photometric data are not 
significantly affected by the choice of IMF.
}
{An} example of the resulting CMD is shown in Fig\,\ref{fig:dSphCut}.
To build 50 mock data for each dSph, the whole
process is repeated 50 times by adding a Gaussian
noise consistent with the uncertainty in the number of
member stars estimated by \citet{2008ApJ...684.1075M}.
The position and velocity of each star are assigned {consistently} with 
the input dark matter potential using the method of \citet{1991MNRAS.253..414C} with the assumptions of 
the constant velocity anisotropy and spherical distribution. 

{The} non-member stars belonging to the Milky Way galaxy are also included,
which are generated by the Besan\c{c}on model\,\citep{Robin:2004qd}
{as shown by blue points in Fig\,\ref{fig:dSphCut}}.%
\footnote{{
We have not included photometric errors for foreground stars, 
for the number of the foreground stars above the limiting magnitudes is less sensitive to the inclusion of the photometric errors than the case of the dSph member stars.
}}
{
The Besan\c{c}on model generates a set of stars
without positional information of individual stars for a given direction.
In this paper, we have used the same set of the foreground stars (generated by the Besan\c{c}on model) for all the 50 mock data, while generated the random distributions of the stars on the sky plane in individual mocks.
}
{We note that an appearent sharp cut-off at $(g-i)<0.3$ seen
only for the foreground stars reflects different assumptions about
the stellar evolution for horizontal branch stars, which
is highly model dependent. Fraction of such blue stars in both the
mock member and foreground stars are
minor compared to red-giant stars or the main-sequence stars
in the mock data.
}
\subsection{Spectrograph}

In our analysis, we adopt the same detector capability in \KI (see Table\,3 in \KI).
The observing parameters  are based on the capability of the 
Prime Focus Spectrograph (PFS) {attached to} 8.2\,m Subaru telescope.
PFS is the next generation spectrograph of the SuMiRe project\,\citep{2014PASJ...66R...1T, 2015JATIS...1c5001S,Tamura:2016wsg}
and the science operation is planned {to start around} $2019 \-- 2020$.
The key advantages of PFS are its large field-of-view ({$\sim 1.38^{\circ}$ diameter}), 2394 fibers,
and the wide wavelength {coverage ($380 \-- 1260$ nm) mounted on the large aperture telescope}.
{
One of the main targets is the classical dSphs (Fornax, Sculptor, Draco, Ursa\,Minor, and Sextans), 
for which line-of-sight velocities of stars are measured with a precision dv of $\sim 3 \unit{km/s}$ 
down to magnitudes deeper than $i \sim 21$ covering a wide area well beyond their tidal radii.
}
{
The unique capability of PFS has also an advantage in observing ultra-faint dwarf galaxies, increasing the sample size by a factor of 2 or more and simultaneously covering the target galaxy and the foreground/background Milky Way stars. The latter aspect is crucial in efficiently taking the effect of contaminating stars into account as in the analysis presented later in this paper. 
}

To take the spectroscopic capabilities into account, 
we smear the 
{
mock velocity, surface gravity ($\log g$) and metallicity ([Fe/H]) data with widths corresponding to the expected measurement errors, 3\,km/s, 0.5\,dex and 0.5\,dex, respectively
} 
and 
select the stars which locate {at} $r < d \sin \thetaROI$, reflecting the limitation of the region of interest.
Here $d$ denotes the distance of each dSph and $\thetaROI$ is the angular radius of the region of interest.

The depth of the survey depends on the exposure time.
We adopt three cases of the upper bound of 
the magnitude ($\imax = 21,\,21.5$ and $22$).
In the first case, we demonstrate  the current sensitivity reach.\footnote{
Since the size of the {UFD}s is smaller than that of the classical dSphs,
the kinematical data provided by the current observations is deeper than
the classical dSphs ($i \sim 19.5 $).
}
The second case ($\imax = 21.5$) is for a deeper survey with an integration time of several nights.
The third case is for an ultimate reach.

\begin{table*}
\begin{center}
  \begin{tabular}{ccccccc}
  {Model} dSph
& $v_{\text{dSph}}$ [km/s] 
& {$r_\text{max}$ [pc]}
& $ \log_{10} (g/[\text{cm}/\text{s}^2])_{\text{lower}} $
& $ \log_{10} (g/[\text{cm}/\text{s}^2])_{\text{upper}} $
& $[\text{Fe/H}]_{\text{lower}}$
& $[\text{Fe/H}]_{\text{upper}}$
\\ \hline
Ursa\,Major\,II 	& -116.5 	& {294}	& $0.2$ 	& $4.9$ 	& $-4.5$ & $-1.5$ \\
Coma\,Berenices	& 98.1 		& {238}	& $0.1$ 	& $4.7$ 	& $-4.3$ & $-1.5$ \\
Segue\,1 			& 208.5 	& {139}	& $0.9$ 	& $5.1$ 	& $-6.1$ & $-1.2$ \\
Ursa\,Major\,I 	& -55.3 	& {732}	& $0.0$ 	& $3.7$ 	& $-4.9$ & $-0.9$ \\ \hline
 \end{tabular}
\caption{\sl
The bulk velocity, the truncation radius and cut conditions for each dSph.
The bulk velocity of Ursa\,Major\,II, Coma\,Berenices, and Ursa\,Major\,I is from 
\citet{Simon:2007dq} and Segue\,1 from \citet{Simon:2010ek}.
{The truncation radii are from \citet{Geringer-Sameth:2014yza}.}
See the text for {more details}.
}
\label{tb:Cut}
\end{center}
\end{table*}

\begin{table*}
\begin{center}
  \begin{tabular}{c@{\hspace{3em}}cc@{\hspace{3em}}cc@{\hspace{3em}}cc@{\hspace{3em}}cc}
&\multicolumn{2}{c}{\hspace{-3em}Condition}
& \multicolumn{2}{c}{\hspace{-3em}Raw}
& \multicolumn{2}{c}{\hspace{-3em}Naive cut}
& \multicolumn{2}{c}{\hspace{-2em}Membership selection}
 \\ \hline
  {Model} dSph
& $\thetaROI$ [degree] 
& $\imax$ [mag] 
& $N_{\text{Mem}}$
& $N_{\text{FG}}$
& $N_{\text{Mem}}$
& $N_{\text{FG}}$
& $N_{\text{Mem}}$
& $N_{\text{FG}}$
\\ \hline
Ursa\,Major\,II
& 0.65  & 21 & 80 & 829 & 76 & 75 & 54 & 5 \\ 
&        & 21.5 & 150 & 988 & 141 & 103 & 89 & 4 \\ 
&        & 22 & 233 & 1149 & 214 & 132 & 131 & 4 \\ \hline
Coma\,Berenices
& 0.65 & 21 & 35 & 579 & 34 & 58 & 29 & 2 \\ 
&        & 21.5 & 58 & 743 & 55 & 85 & 44 & 2 \\ 
&        & 22 & 92 & 898 & 85 & 110 & 66 & 1 \\ \hline
Segue\,1 
& 0.65 & 21 & \red{24} & \red{620} & \red{22} & \red{60} & \red{19} & \red{1} \\ 
&        & 21.5 & \red{43} & \red{748} & \red{39} & \red{84} & \red{34} & \red{0} \\ 
&        & 22 & \red{61} & \red{953} & \red{56} & \red{123} & \red{49} & \red{0} \\ \hline
Ursa\,Major\,I
& 0.65 & 21 & 42 & 680 & 37 & 32 & 26 & 1 \\ 
&        & 21.5 & 55 & 831 & 48 & 39 & 34 & 1 \\ 
&        & 22 & 63 & 953 & 56 & 44 & 38 & 1 \\ \hline
 \end{tabular}
\caption{\sl
The averaged numbers of the member (foreground) stars
are given by $N_{\text{Mem}}$ ($N_{\text{FG}}$).
{
The {\sl Raw} column shows
the numbers of the stars after the colour-magnitude cut and the cut of the region of interest.
The details of the naive cuts and membership selection are given in the text.
}
}
\label{tb:Ns}
\end{center}
\end{table*}

\subsection{Data selection}

\begin{figure}
\centering
\includegraphics[width=70mm]{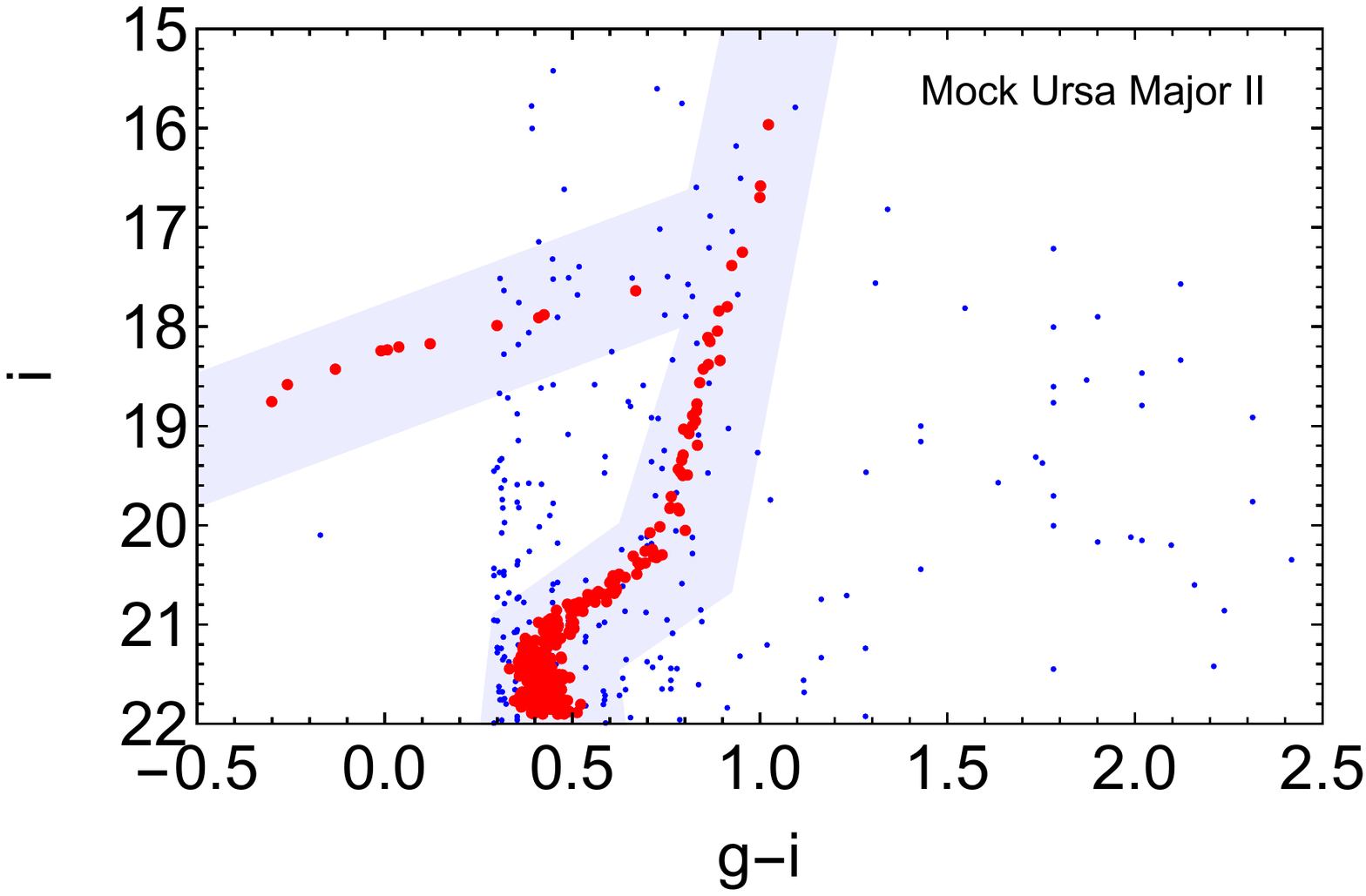} \\
\includegraphics[width=70mm]{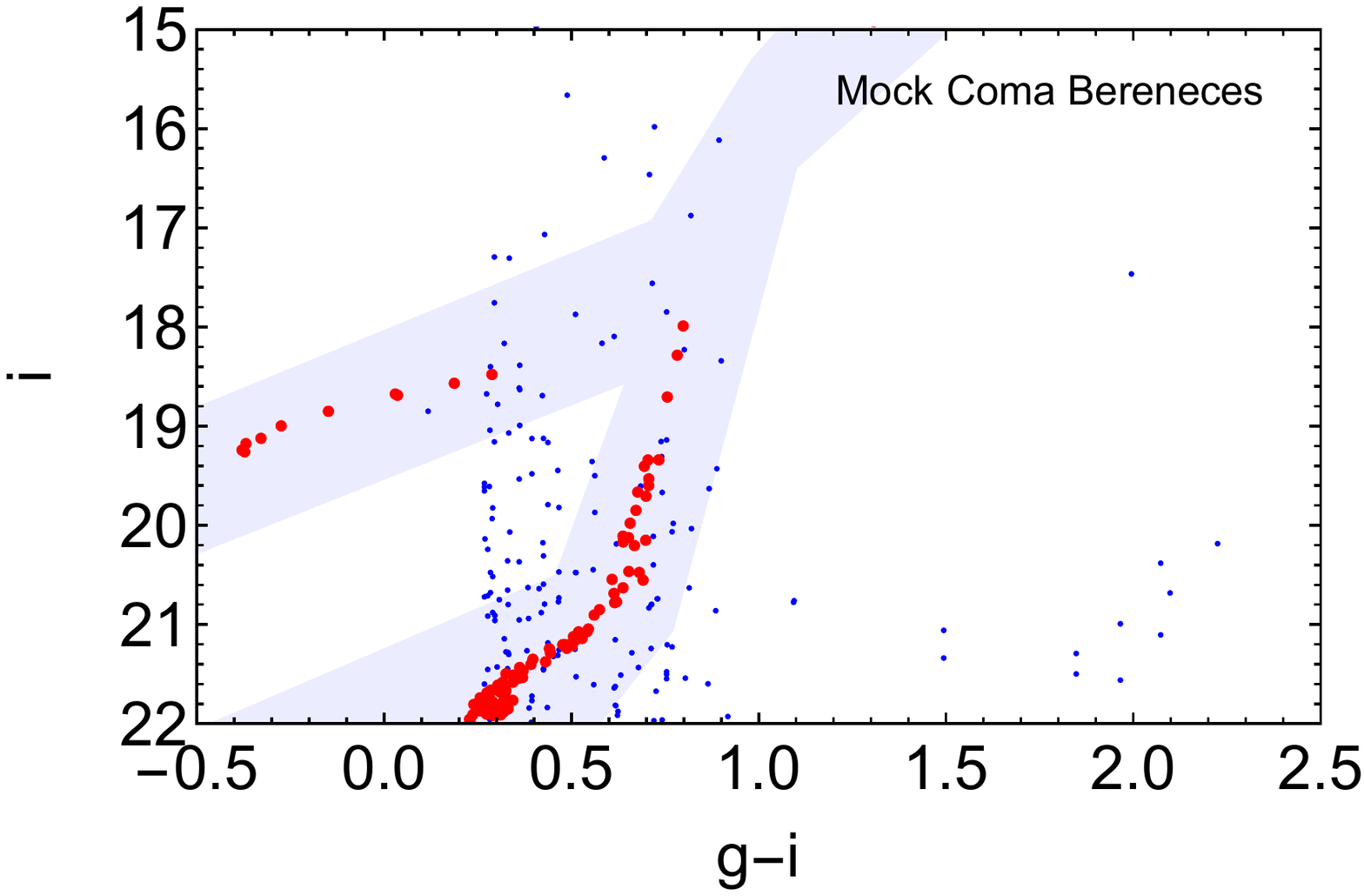} \\
\includegraphics[width=70mm]{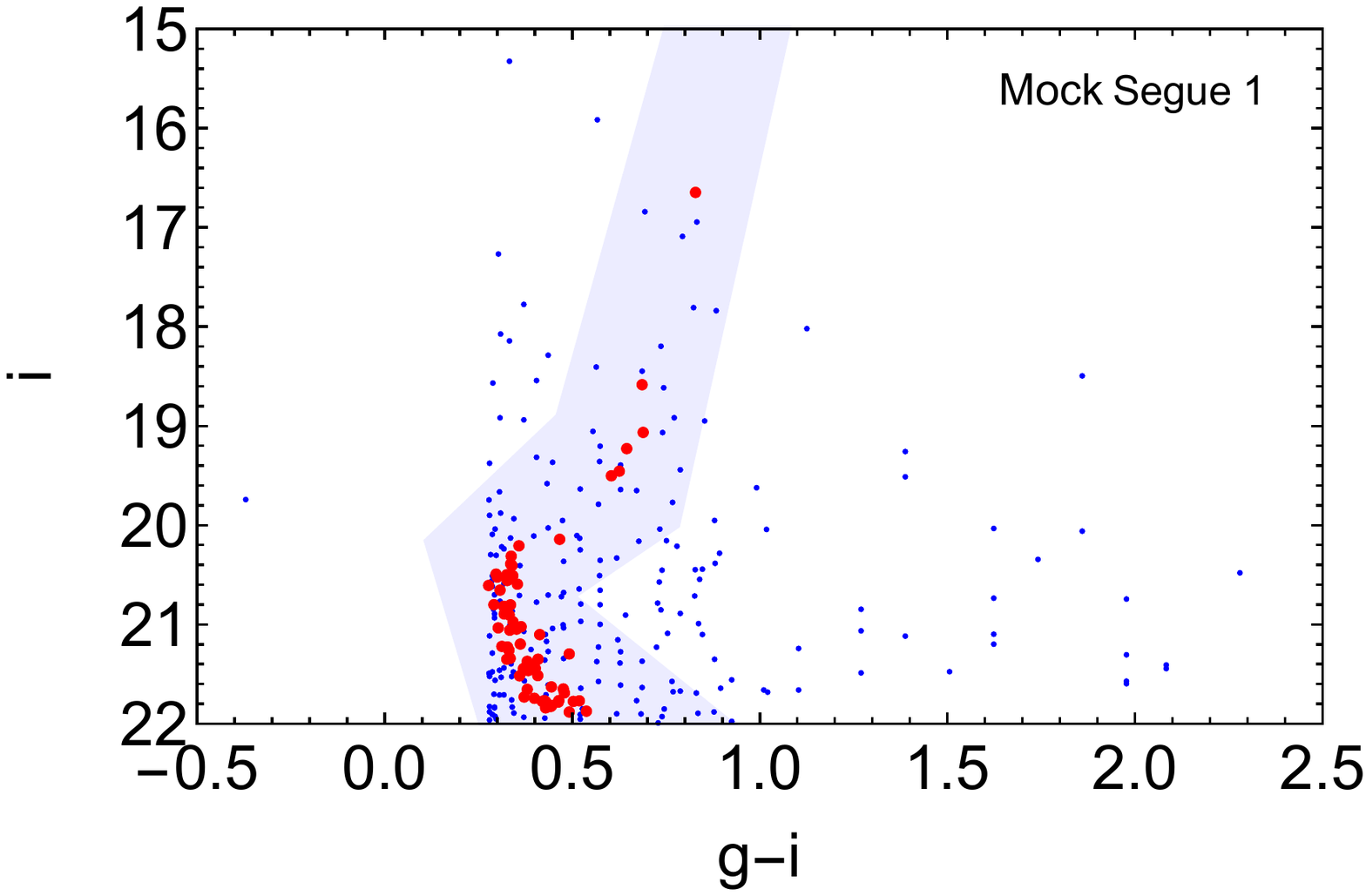} \\
\includegraphics[width=70mm]{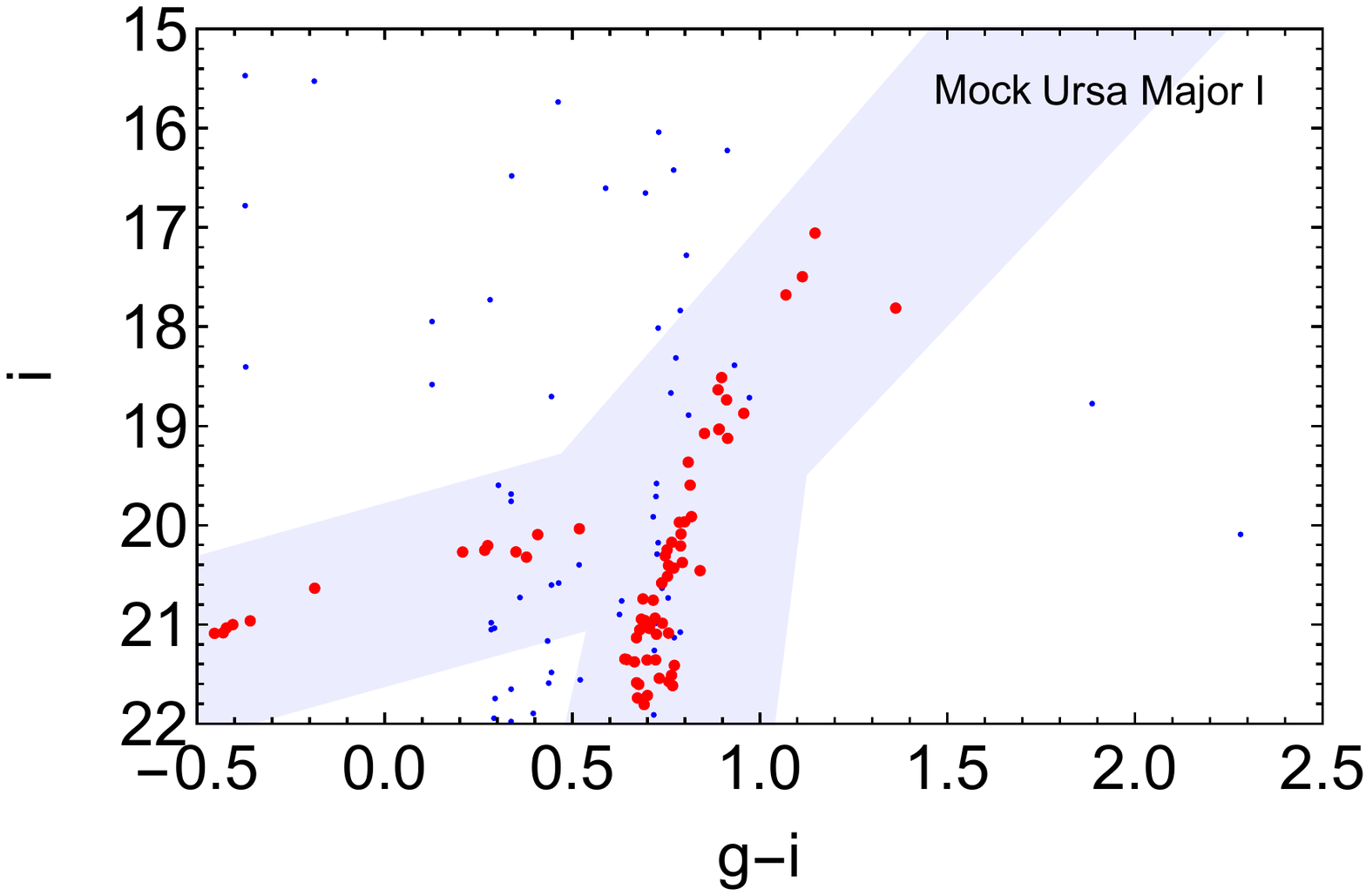}
\caption{\small\sl 
The colour-magnitude map for each dSph.
We impose the colour-magnitude cut by the blue shaded region.
The red (blue) dots show the members (foreground) stars.
The stars on the map are residuals after the cuts of the ROI, velocity, and $\log g$.
}
\label{fig:dSphCut}
\end{figure}

Before the likelihood analysis, 
the foreground contamination in the mock data 
can be largely reduced 
by using the information of its position, velocity, surface gravity, metallicity, and colour-magnitude.
We here adopt two approaches {to} the data reduction: 
naive cut approach and more sophisticated membership selection.

\subsubsection{Naive cut}
In this approach,
we impose the cuts of the velocity,  surface gravity, metallicity, and colour-magnitude
on the dataset and optimize them by (roughly) tuning the boundaries of the cuts by eye.
The velocity cut is a $\pm 60$ [km/s] range from each bulk velocity $v_{\text{dSph}}$. 
The lower and upper bounds of the surface gravity $g$ and {metallicity} [Fe/H] are given in Table~\ref{tb:Cut} for each dSph, 
while the region of the colour-magnitude diagram is shown in Fig.\,\ref{fig:dSphCut}.
Note that we choose these boundaries to include most of the stars in clumps.
Although harder cuts can be imposed to reduce the fraction of the foreground stars,
the cut eliminates scattered member stars and
the reconstructed velocity distribution can be distorted and derive a bias of the halo estimation.
We provide the numbers of the member and foreground stars after the cuts 
in the `{\sl Naive cut}' column in Table\,\ref{tb:Ns}.

\subsubsection{Membership selection}
The latter strategy utilizes the membership probability of each star.
The membership probability is defined by the probability 
to find a member star at a given position, velocity, surface gravity, and, metallicity.
We calculate this membership probability 
by a conventional approach given by \citet{Walker:2008fc}.
In the calculation, the distributions of the foreground stars are also taken into account.
The distributions of the velocity, surface gravity, and metallicity 
except for the foreground velocity distribution
are modeled by 
$R$-independent single Gaussians,
while the foreground velocity distribution is fixed without free parameters
and 
the {spatial} distributions are more generally {parametrized}.
The detail of this process is given in \citet{Walker:2008fc} and the appendix of \KI.
We select stars within $95$\% confidence level of the membership probability.
We provide the numbers of the member and foreground stars after this selection
in the `{\sl Membership selection}' column in Table\,\ref{tb:Ns}.
Compared with the case of the `{\sl Naive cut}', 
much higher purities of the data is obtained by this procedure,
while some fraction of the member star is eliminated.
We here stress that this approach assumes the constant velocity dispersion 
in the membership probability assignment.
The member stars eliminated in the selection are mostly due to the constant velocity-dispersion bias and 
therefore \red{can} affect the estimation of the $J$-factors.


\subsection{Kinematical fit}
\label{subsec:Kinfit}

\begin{figure}
\centering
\includegraphics[scale=0.37]{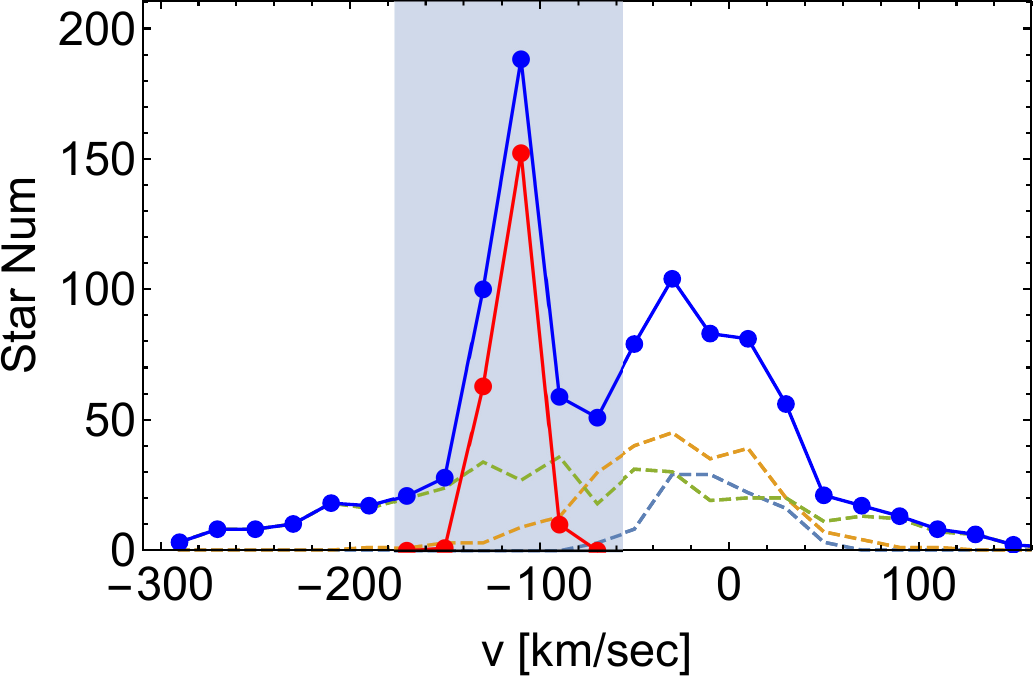}
\caption{\small\sl 
{The velocity distributions of a mock sample of the foreground and dSph (mock \UII) member stars. 
The red solid line shows the distribution of the dSph member stars. The blue solid line shows the sum of the distributions of the dSph and foreground stars.
Dashed blue, orange and green lines correspond to the three components of the foreground stars (thin disk, thick disk and halo). 
\red{The shaded range is the signal region of the velocity $v_\textrm{lower} < v < v_\textrm{upper}$.}
}
}
\label{fig:vDist}
\end{figure}

In this section, we provide two types of the analysis for the kinematical fit.
The first one is the single component fit,
in which all the data is regarded as member star.
The second one is the mixed component approach  developed in \KI, 
in which the member and foreground distribution are simultaneously fitted.
For both the fits, we apply the unbinned likelihood analysis to the halo estimation.

\subsubsection{Single Component fit}
The single component fit is performed by assuming 
that the data 
used for the fit contains only member stars,
which implies
that the likelihood function is given by
\begin{eqnarray}
-2 \ln \Ls = -2 \sum_{i} \ln ( f_{\text{Mem}} (v_{i}, R_{i}))\ ,
\label{eq:likeli_single}
\end{eqnarray}
where
$f_{\text{Mem}}(v,R)$ is the distribution function of the member stars.
The index $i$ runs all the stars in the mock data set.
We assume that the velocity distributions of the member stars
can be  approximated by a single Gaussian and hence the 
distribution functions can be expressed as
\begin{eqnarray}
\hspace{5pt} f_{\text{Mem}}(v,R) = 2 \pi R \Sigma_{*}(R)\,C_{\text{Mem}}\,{\cal G}[v;\,v_{\text{Mem}},\,\sigma_{l.o.s} (R)]\ .
\end{eqnarray}
Here ${\cal G}[x;\,\mu,\sigma]$ denotes the Gaussian distribution of a variable $x$ with a mean value $\mu$ and a standard deviation $\sigma$.
We note that the parameter $v_{\text{Mem}}$ represents the bulk velocity of the dSph and mostly converges to the input bulk velocity $v_{\text{dSph}}$.
The distribution functions are normalized by $C_{\text{Mem}}$ to satisfy
$\int^{r_{\text{ROI}}}_{0}dR \int^{v_{\text{upper}}}_{v_{\text{lower}}} dv f_{\text{Mem}}(v,R) = 1$
where $r_{\text{ROI}} \equiv d\,\sin \thetaROI$.

\subsubsection{Mixed Component fit}
\label{subsubsec:MCfit}
In the mixed component fit, 
the stellar distribution is considered to be the sum of the
foreground and member star distribution.
The likelihood function $\Lm$ is defined by 
introducing 
the membership fraction parameter $s$ as follows
\begin{eqnarray}
-2 \ln \Lm = -2 \sum_{i} \ln ( s f_{\text{Mem}} (v_{i}, R_{i}) + (1-s) f_{\text{FG}} (v_{i}, R_{i}) )\ ,
\label{eq:likeli}
\end{eqnarray}
where
$f_{\text{FG}}(v,R)$ is the distribution function of the foreground stars.
We model the foreground distribution function 
by the production of the three Gaussians,
corresponding to the foreground thin disc, thick disc, and halo components:
\begin{eqnarray}
f_{\text{FG}}(v,R) = 2 \pi R\,C_{\text{FG}} \prod_{j=1}^{3} {\cal G}[v;\,\vFGj,\,\sigmaFGj]\,,
\end{eqnarray}
with $\vFGj$, $\sigmaFGj$ ($j = 1,\,2,\,3$) 
being parameters of the distribution.
Here we assume that the parameters $\sigmaFGj$ are independent of $R$ in contrast to the dispersion of the member star.
The constant $C_{\text{FG}}$ denotes the normalization factor to satisfy
$\int^{r_{\text{ROI}}}_{0}dR \int^{v_{\text{upper}}}_{v_{\text{lower}}} dv f_{\text{FG}}(v,R) = 1$.

For the sake of the convergence of the mixed component fit,
we constrain the parameters $\vFGj$, and $\sigmaFGj$
by using the data in the control region
(i.e., the region in which the number of the member stars is negligible).
In \KI, we deduce the foreground velocity distribution
by using the data out of the region of the velocity cut 
and interpolate it to the signal region.
For the UFD case, on the other hand, 
since the bulk velocities of these dSphs are
not as large as that of classical dSphs,
the foreground estimation by using the control region in the velocity distribution
does not efficiently work.
{In the case of the classical dSphs considered in \KI, the bulk velocity of the member stars is largely different from the Milky Way stellar halo component, which makes the velocity cut work most efficiently. As Fig.\,\ref{fig:vDist} shows, on the contrary, the distribution of the foreground stars overlaps with the distribution peak of the UFD, hence the significant amount of the foreground stars contaminate the signal region, which prevent using the same method in \KI. }

Instead, 
we define the control regions  in the distribution of the spatial position
by setting an annulus centred at each dSph galaxy 
from the radius of the signal region to the PFS threshold, $\theta = 0.65^\circ$.
{
For precise determination of the DM profile and increasing the value of the $J$-factor,
we need to take a large signal region as possible,
though the signal region must be included in PFS threshold 
and the range for the control region must be reserved.
}
Here, the radii of the signal regions are chosen to be
$2 r_{e},\,4 r_{e},\,4 r_{e}$, and $r_{e}$ 
for \UII, \Coma, \Seg, and \UI respectively,
based on their half-light radii $r_e$.
{
We should bear in mind that the radius of the signal region is not optimized. 
Instead, we tried to take the radius so that its angular diameter is about a half of FoV.}

When we perform a fit to the control region,
we take into account 
the effect of the thin and thick disc components of the foreground stars in addition to the halo component,
because the disc components remain after surface gravity and metallicity cuts in case of UFDs.
This contrasts to the case of classical dSphs, 
where the foreground stars mainly belong to the halo component after the naive cut.
In order to represent the three foreground components 
we assume the foreground distribution can be expressed by a sum of three Gaussian functions.
We first perform fits by the three Gaussian model for control region data
on which colour-magnitude, $\imax$ and ROI cuts are imposed,
and obtain the best-fitting values and standard deviations of each Gaussian. 
Then we perform secondary fits for the control region data with all naive cuts
(colour-magnitude, $\imax$, ROI, surface gravity and [Fe/H]) imposed on, 
using the best-fitting Gaussians achieved in the first fit as the priors.
Here we obtain the best-fitting values and standard deviations of $\vFGj$, 
$\sigmaFGj$, which are defined as
$\vFGjo$, $\sigmaFGjo$, $d\vFGj$, and $d\sigmaFGj$ respectively. 
Finally we use this information as a prior for $\vFGj$, $\sigmaFGj$ 
by multiplying 
$
\prod_{j = 1,2,3}
{\cal G}[\vFGj;\,\vFGjo,\,d\vFGj]\,{\cal G}[\sigmaFGj;\,\sigmaFGjo,\,d\sigmaFGj]$
to the likelihood function ${\cal L}$ in Eq.(\ref{eq:likeli}).

\subsubsection{Fit algorithm}
The likelihood function (multiplied by the foreground priors for the mixed component fit) are searched 
by performing the Metropolis-Hastings algorithm\,\citep{Metropolis:1953am,Hastings:1970aa} of the Markov Chain Monte Carlo (MCMC) method.
The parameter set of the single component fit consists of 
the five free parameters of the dark matter halo
($\rho_s,\,r_s,\,\alpha,\,\beta,\,\gamma$), 
one velocity anisotropy parameter $\beta_{\text{ani}}$
and 
one nuisance parameters ($v_{\text{Mem}}$),
while the mixed component fit also has 
the other seven nuisance parameters 
($s$, $\,\vFGj$, $\sigmaFGj$).
In the MCMC method, the halo parameters are searched under the flat/log-flat priors within the range of
$-4 < \log_{10} (\rho_{s}/[\text{M}_{\bigodot}/\text{pc}^3]) < 4 $, 
$ -2 < \log_{10} (r_{s}/[\text{kpc}]) < 5 $,
$0.5 < \alpha < 3  $,
$ 3 < \beta < 10  $,
$ 0 < \gamma < 1.2  $ and 
$-1 < \log_{10} (1-\beta_{\text{ani}}) < 1 $.

\subsection{Strategy}

Using 50 mocks for each case ($\imax = 21,\,21.5$, and 22),
we test three types of the $J$-factor estimation:
the method of \KI (naive cut + mixed component fit),
\MS analysis (membership selection + single component fit),
and \Cont fit (naive cut + single component fit).
We here stress that 
in the \KI  approach,
the velocity distribution of the foreground is parametrized by the fit
and therefore the error bar of the $J$-factor involves the uncertainty of the foreground distribution,
while we fix the spatial stellar distributions of member and foreground stars in the likelihood.
This contrasts with the \MS approach in which 
a fixed model of the foreground velocity distribution and 
parametrized spatial distributions are used in the selection.

\section{Results}
\label{sec:results}
\begin{figure*}
\begin{center}
\includegraphics[width=120mm,angle=0]{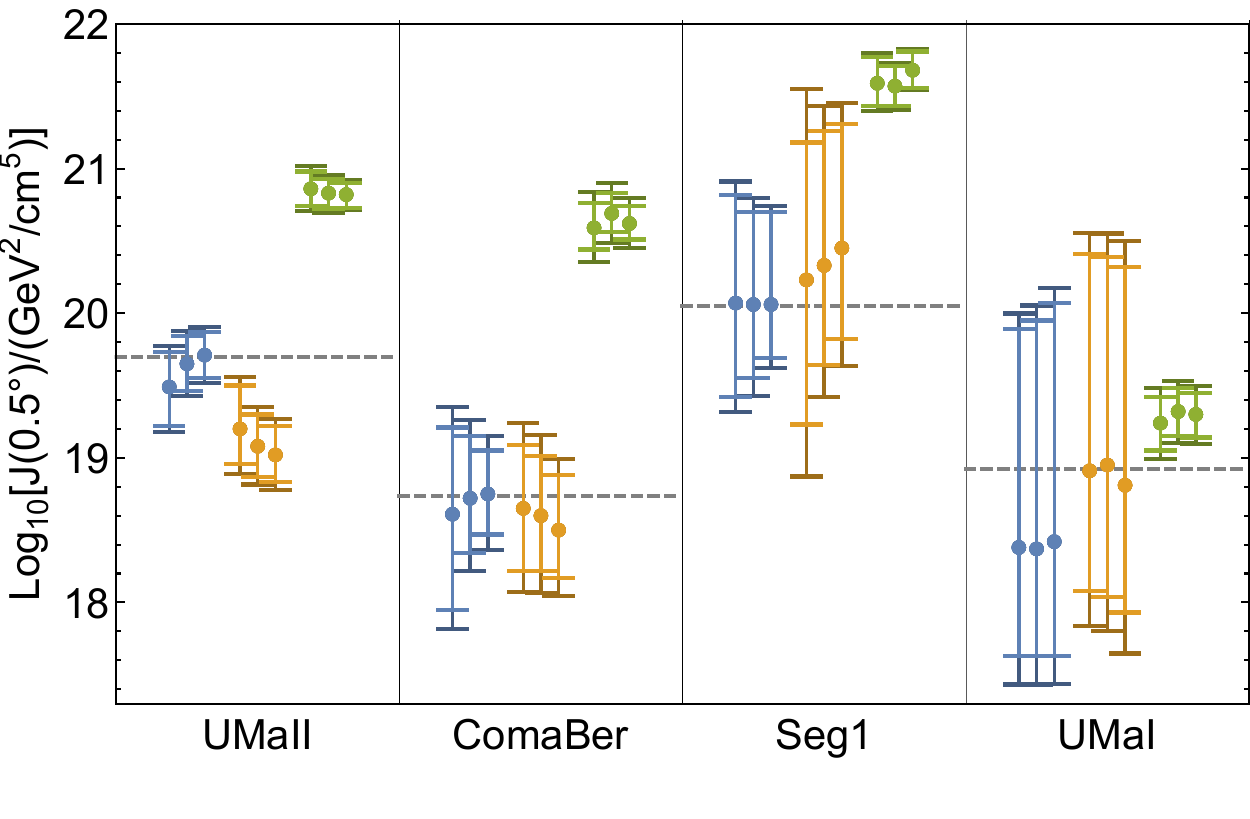}
\vspace{15pt}
\caption{\small\sl
The $J$-factors obtained by the fits are plotted. 
The blue, orange, and green dots show the $J$-factor estimation{s} of  \KI, \MS and \Cont analysis. 
The lighter error bars of each point
show the average of the $68\,\%$ quantile, 
while the darker ones show the square root of the $68\,\%$ quantiles and the standard deviation of the median values.
The grey dashed lines show the input values. 
For each dSph, 
three bars with the same colours 
correspond to the case of $\imax = 21,\,21.5$, and $22$ with $\thetaROI = 0.65$
respectively, from the left.
{See Table\,\ref{tab:Jtable} \red{and Figure \ref{fig:trunc}} for the numerical values \red{and their dependence on the $r_\text{max}$ , respectively}.}
}
 \label{fig:Jfig}
 \end{center}
\end{figure*}

Fig.\,\ref{fig:Jfig} shows the results of these three approaches, 
namely, the method of KI17, the Conventional analysis, and the Contaminated fit
by blue, orange, and green bars, respectively.
Here we give the averaged median values of $\log_{10} (J/[\text{GeV}^2/\text{cm}^5])$ for each fit by the dots.
The lighter error bars show the averages of the widths of the $68\,\%$ quantiles, 
while the darker ones show the square roots of the $68\,\%$ quantiles and the standard deviations of the median values,
written in an additional way to the lighter ones.
The grey dashed lines show the input values. 
For each dSph, 
three bars with the same colours
correspond to the case of $\imax = 21,\,21.5$, and $22$ with $ \thetaROI= 0.65$ 
respectively, from the left.
All $J$-factors are calculated within 
an angular radius of $0.5 \,\text{degree}$ 
(i.e., $\Delta \Omega = 2.4 \times 10^{-4}\,\text{sr}$),
which is the standard size for the $J$-factor calculation.
We here choose 
{the distance from the centre of the dSph to the outermost observed member star $r_\text{max}$ as the most conservative radius}
given by \,\citet{Geringer-Sameth:2014yza}.


In the \Cont analysis (green bars in Fig \ref{fig:Jfig}),
the overestimation of the $J$-factor becomes more than an order of the magnitude.
This is because the dispersion curve inflates
due to the foreground contamination 
which mainly locate at the outer region with a large velocity dispersion ($\sim 30 \-- 40 $ km/s).
Since the fraction of the foreground {contaminating stars} is more than 50\,\%,
the overestimation is much larger than that of the classical case.

\begin{figure*}
 \begin{center}
  \includegraphics[width=60mm,angle=0]{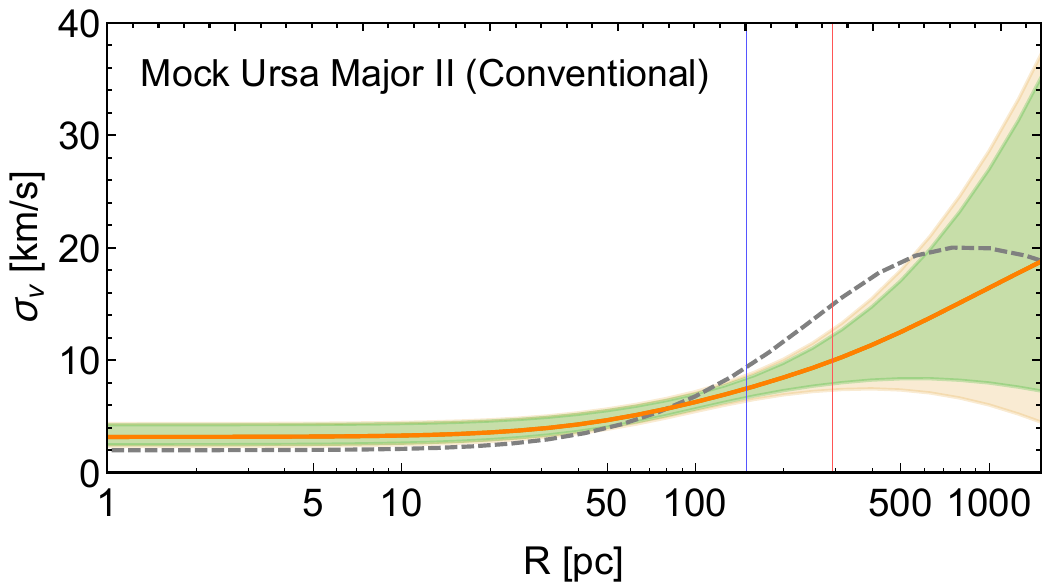}\hspace{1em}
  \includegraphics[width=60mm,angle=0]{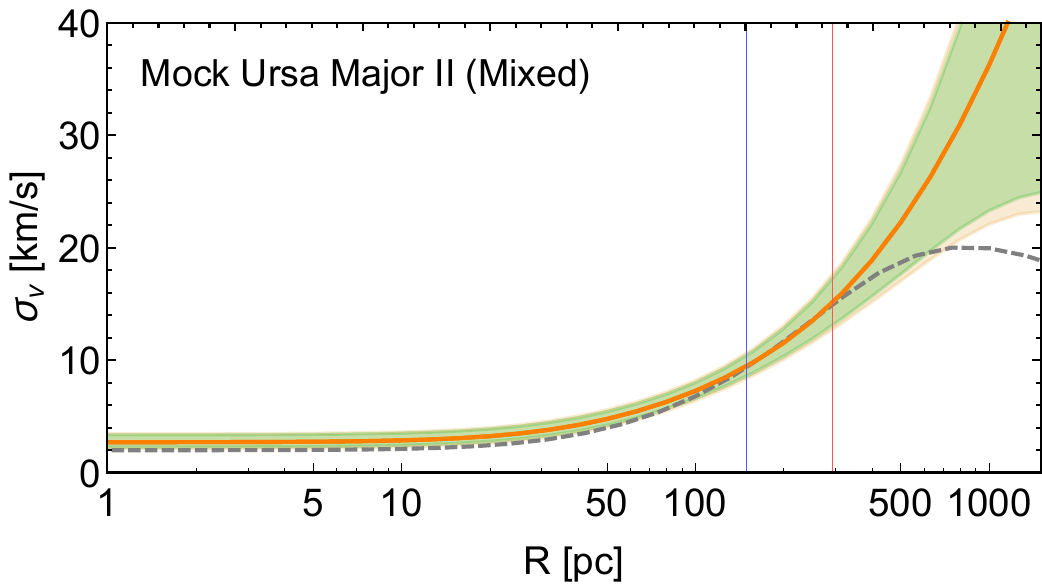} 
\\
  \includegraphics[width=60mm,angle=0]{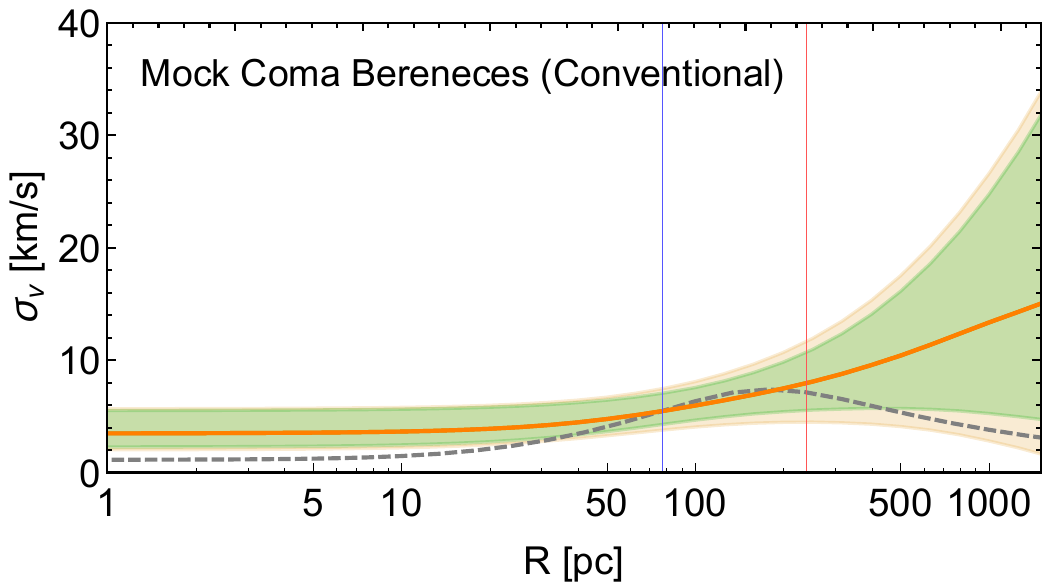}\hspace{1em}
  \includegraphics[width=60mm,angle=0]{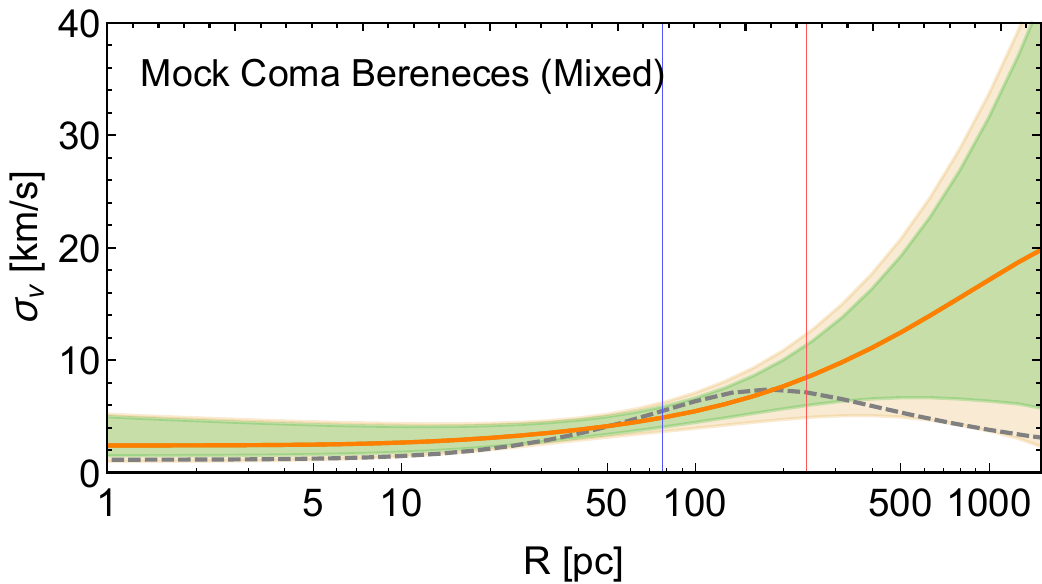} 
\\
  \includegraphics[width=60mm,angle=0]{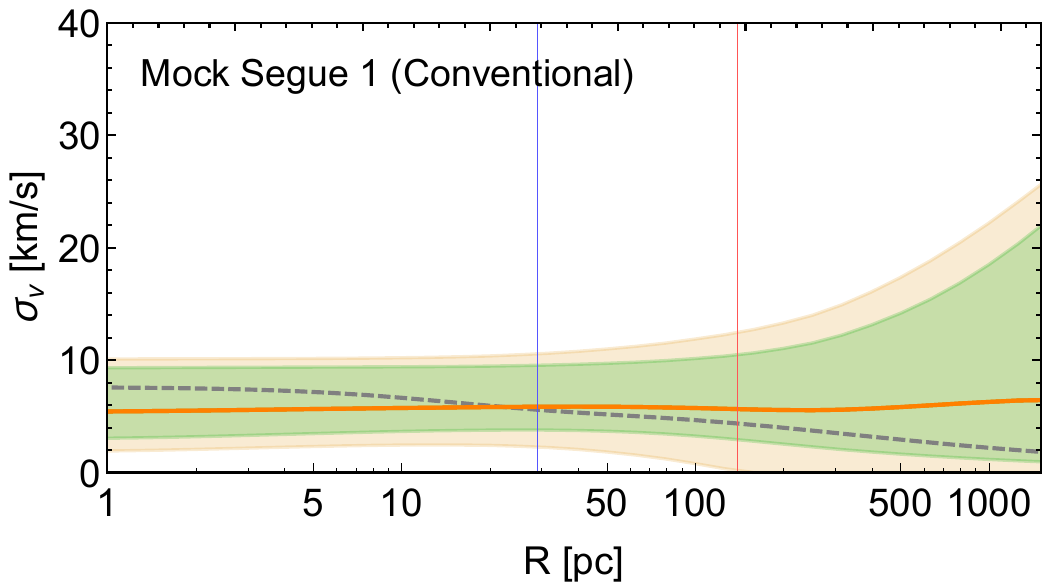}\hspace{1em}
  \includegraphics[width=60mm,angle=0]{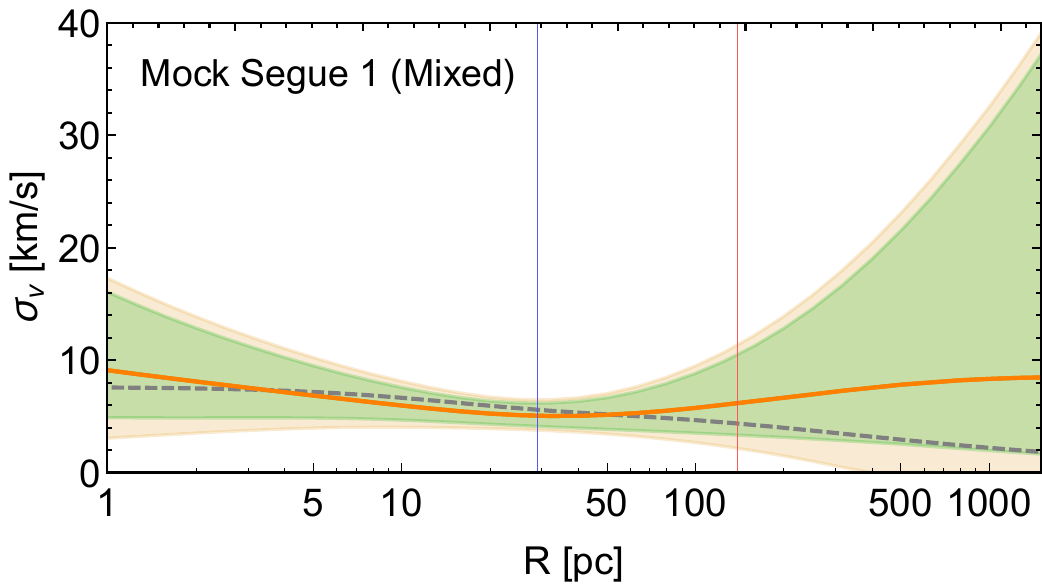} 
\\
  \includegraphics[width=60mm,angle=0]{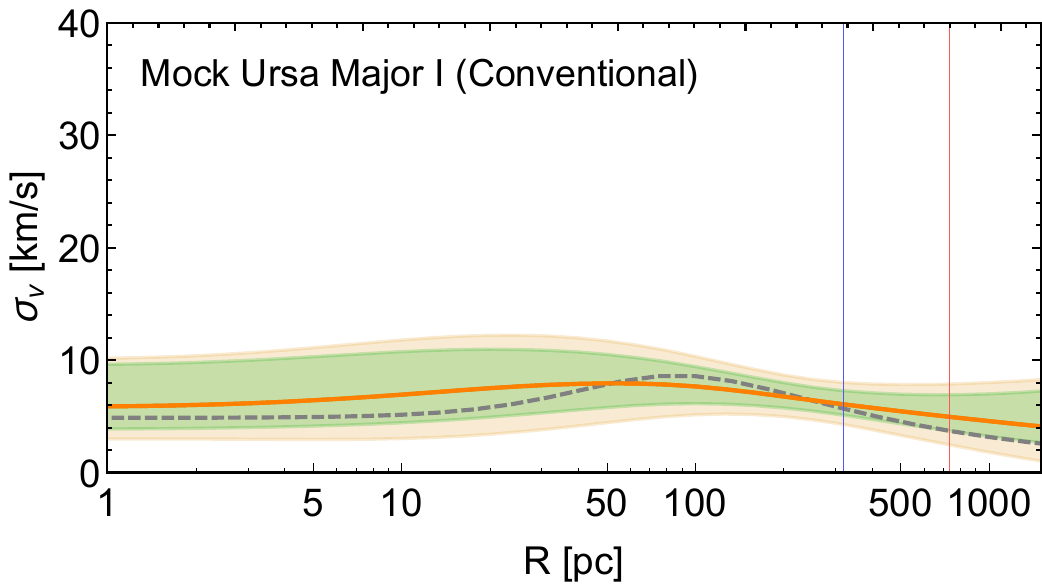}\hspace{1em}
  \includegraphics[width=60mm,angle=0]{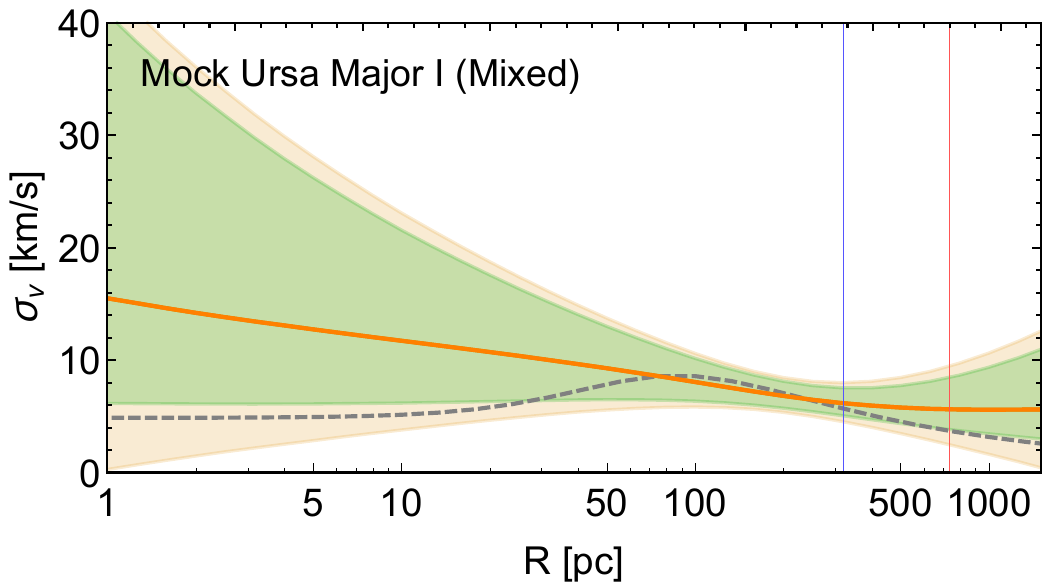} 
 \end{center}
 \caption{
\small \sl
{\bf Left panel:} 
The distribution of the dispersion curve obtained by the fit of the \MS approach.
The red lines show the median value of the dispersion curves 
(averaged by the 50 mocks).
The green bands show the (averaged) 68\,\% quantile.
The median value of the dispersion curve also fluctuate sample by sample,
reflecting the quality of the sample.
The fluctuations by the quality of the samples are shown  by the orange shaded regions,
which are obtained by 
the square root sum of the standard deviation of the median values of the 50 mocks
and the 68\,\% quantiles. 
The input dispersion curves are also shown by the grey dashed lines.
\red{The vertical blue(red) lines correspond to $r_e$($r_\text{max}$). }
 {\bf Right panels:}
The same figures as the left panels
but obtained by the fit of the \KI approach.
}
 \label{fig:EMMFdisp}
\end{figure*}

On the other hand, 
the $J$-factors seem to be successfully reproduced by the \MS approach 
(orange bars in Fig \ref{fig:Jfig}).
However, since this approach assumes a constant velocity dispersion in its membership \red{selection}, 
the dispersion curve after the selection tends to be more or less constant as a function of the radius,
which lead to a small bias to the $J$-factor estimation.
We provide the typical uncertainty in the dispersion curve and its sample-to-sample scatter for the 50 mocks {in the left column of Fig.\,\ref{fig:EMMFdisp}}.
The red lines show the median value of the dispersion curves 
obtained by the fit of the \MS approach (averaged by the 50 mocks),
while the green band shows the (averaged) 68\,\% quantile.
The median values of the dispersion curves also fluctuate sample by sample,
reflecting the quality of the sample.
We show this  fluctuation by the orange shaded regions
which are obtained by 
the square root sum of the standard deviation of the median values of the 50 mocks and the 68\,\% quantiles.
The input dispersion curves are also shown by the grey dashed lines.

\red{For the \UII, the figure shows that the dispersion curve is flatter than that of the input, 
which makes the $J$-factor underestimated.
This fact seems to be caused by the constant velocity dispersion bias.
This effect becomes more significant for a larger size of stellar data, as can be seen in the three orange bars.}
Meanwhile, since the changes of the dispersions curves of the other dSphs 
are not as  large as the \UII case,
the \red{effect of the bias is not seen}.
\vspace{0.1cm} 

We also note the results for the \UI case.
Although the number of the stars in \UI does not significantly 
differ from the other dSphs (see Table\,\ref{tb:Ns}),   
both the {\MS} and \KI approaches cannot determine the $J$-factor 
as precisely as those of the other dSphs.
It seems to originate in the fact that the inner part of the dark matter profile is not well determined.
We left this analysis to future work.


\KI approach (blue bars in Fig. \ref{fig:Jfig}) also provides successful $J$-factor estimations.
\red{For the \UII case, the $J$-factor  is getting converged to the input value for deeper observations, 
while it cannot be seen in the conventional approach.}
\red{For other dSphs, the result of our analysis is compatible with the conventional one,}
\red{which reflects the fact that the velocity dispersion at the inner part is more or less} \red{constant.}

We \red{also} give the distribution of the dispersion curve obtained by the fit {in the right column of Fig.\,\ref{fig:EMMFdisp}}.
Compared with the distribution of the \red{\MS} approach, 
the width of the 68\,\% quantile is larger in the outer region, 
while it becomes smaller in the inner region (except for the \UI case).
The results of the $J$-factor estimation implies that the width in the inner region
preferentially affects the uncertainties of the $J$-factors.
We also note that the median dispersion curve of the \UII case 
successfully follows the input curve at $R \sim 250$ pc,
in contrast with that of the \red{\MS} approach.
\red{The sensitivity of the indirect dark matter detection observing gamma-rays from the dSphs 
depends directly on the median values and uncertainties of $J$-factors. 
See Appendix \ref{sec:CTAanalysis} for those who are interested in this dependence in a concrete example.
}

\section{Summary}
\label{sec:summary}
In this paper, 
we have investigated the effect of the foreground contamination on the estimation of astrophysical factor,
using the mock kinematical data of the four representative ultra-faint dwarf spheroidal galaxies.
This is because we cannot completely distinguish the foreground stars from the dSph's member stars even if imposing several data cuts.
We have adopted our developed fitting analysis, \KI, utilizing the future spectroscopic survey, PFS.
Such a multi-object spectrograph with large field of view enables us 
to observe numerous number of stellar spectra required \red{for} the \KI analysis.

For comparison, we have performed three types of the $J$-factor estimation: the \KI methods, the \MS analysis and the \Cont fit.
As the result of the analysis, 
the $J$-factor value estimated by the \Cont analysis is up to a few hundred times larger than the input value
and its confidence interval is significantly small,
because all stellar data after naive cut are regarded as member star even including the foreground contamination.
On \red{the} other hand, 
the \KI and \MS analysis can reproduce the input $J$-factor value within $1\sigma$ confidence levels except for \UII.

For the case of \UII, \red{which has the non-flat velocity dispersion curve,} the \MS approach underestimates the $J$-factor value with respect to the input value.
\red{This seems to originate from the assumption of the constant velocity dispersion at the membership selection.
Accordingly, the member stars in the outer region tend to be rejected.
}

The likelihood function of the \KI method includes the information of both the foreground stars and the member ones together
with the parameters describing their distribution functions\red{,}
and
the properties of the foreground stars are roughly determined by the photometric and spectroscopic observations of the stars in the control region.
\red{This method allows us} to treat correctly and statistically the effect of the foreground contamination for the observational data.
\red{Moreover, the method can provide the validation of the assumption concerning the velocity dispersion curve in the \MS analysis.}
Therefore, our statistical method should become powerful tool for the $J$-factor estimate of the MW dSphs in the PFS-era\red{.}

\vspace{0.5cm}
\noindent
{\bf Acknowledgments}
\vspace{0.1cm}\\
\noindent
\red{We are grateful to the referee for her/his careful reading of our paper and thoughtful comments.}
We would like to give special thanks to Josh Simon and Marla Geha for giving us the kinematic data of UFD galaxies.
This work is supported in part by 
the Ministry of Education, Culture, Sports, Science, and Technology (MEXT), Japan, 
No. 16H01090 (for K. H.),
No. 15H05889, No. 25105011 (for M. I.), 
No. 16H02176, No. 26104009 (for S. M.), 
No. 17H02878 (for S. M. and M. I.),
No. 17K14249 (for M. N. I.),
\red{and No. 18J21186 (for S. H.).}
The work of K.I. is also supported by the JSPS Research Fellowships for Young Scientists.
Finally, Kavli IPMU is supported by World Premier International Research Center Initiative (WPI), MEXT, Japan.


\bibliographystyle{mnras}
\bibliography{ref3}


\appendix

\section{Table of obtained $J$-factors}
\renewcommand{\arraystretch}{1.4}
\begin{table*}
\centering
\begin{tabular}{cccccc}
 		&		& \red{Mock} \UII 					& \red{Mock} \Coma 					& \red{Mock} \Seg 					& \red{Mock} \UI 					\\
\hline
$i=21 $ 	& \KI 	& \red{\errs{19.49}{0.27}{0.24}{0.15}}	& \red{\errs{18.61}{0.66}{0.60}{0.44}}	& \red{\errs{20.07}{0.65}{0.75}{0.38}}	& \red{\errs{18.38}{0.75}{1.51}{0.58}} \\
	 	& \MS 	& \red{\errs{19.20}{0.24}{0.30}{0.20}}	& \red{\errs{18.65}{0.43}{0.44}{0.39}}	& \red{\errs{20.23}{1.00}{0.95}{0.92}}	& \red{\errs{18.91}{0.83}{1.50}{0.68}}\\
	 	& \Cont 	& \red{\errs{20.86}{0.12}{0.12}{0.10}}	& \red{\errs{20.59}{0.15}{0.17}{0.18}}	& \red{\errs{21.59}{0.16}{0.18}{0.11}}	& \red{\errs{19.24}{0.19}{0.18}{0.16}}\\ \hline
$i=21.5 $ 	& \KI 	& \red{\errs{19.65}{0.19}{0.19}{0.12}}	& \red{\errs{18.72}{0.38}{0.43}{0.33}}	& \red{\errs{20.06}{0.51}{0.64}{0.37}}	& \red{\errs{18.37}{0.74}{1.58}{0.58}}\\
		& \MS 	& \red{\errs{19.08}{0.21}{0.22}{0.16}}	& \red{\errs{18.60}{0.38}{0.41}{0.38}}	& \red{\errs{20.33}{0.69}{0.93}{0.59}}	& \red{\errs{18.95}{0.91}{1.44}{0.70}}\\
	 	& \Cont 	& \red{\errs{20.83}{0.11}{0.10}{0.08}}	& \red{\errs{20.69}{0.13}{0.14}{0.16}}	& \red{\errs{21.57}{0.14}{0.14}{0.08}}	& \red{\errs{19.32}{0.17}{0.16}{0.14}}\\ \hline
$i=22 $ 	& \KI 	& \red{\errs{19.71}{0.16}{0.16}{0.11}}	& \red{\errs{18.75}{0.28}{0.30}{0.27}}	& \red{\errs{20.06}{0.37}{0.64}{0.24}}	& \red{\errs{18.42}{0.79}{1.65}{0.59}}\\
		& \MS 	& \red{\errs{19.02}{0.19}{0.20}{0.15}}	& \red{\errs{18.50}{0.33}{0.38}{0.31}}	& \red{\errs{20.45}{0.63}{0.86}{0.52}}	& \red{\errs{18.81}{0.88}{1.51}{0.76}}\\
	 	& \Cont 	& \red{\errs{20.82}{0.09}{0.08}{0.06}}	& \red{\errs{20.62}{0.11}{0.12}{0.13}}	& \red{\errs{21.68}{0.12}{0.13}{0.07}}	& \red{\errs{19.30}{0.16}{0.15}{0.13}}\\ \hline
\end{tabular}
\caption{{The numerical value of the $J$-factor obtained by our fit, illustrated in Fig.\,\ref{fig:Jfig}.}
\red{We show the averaged median values of $\log_{10}(J/[\text{GeV}^2/\text{cm}^5])$ (the first values), 
the averages of the 68\% quantiles (the first uncertainties) 
and the standard deviations of the median values (the second uncertainties). }}
\label{tab:Jtable}
\end{table*}
\renewcommand{\arraystretch}{1}
We show the numerical values of the $J$-factors and its errors obtained by our fits in Table \ref{tab:Jtable}.

\section{Implication to gamma-ray detections}
\label{sec:CTAanalysis}
\begin{figure*}
\begin{center}
\includegraphics[width=70mm,angle=0]{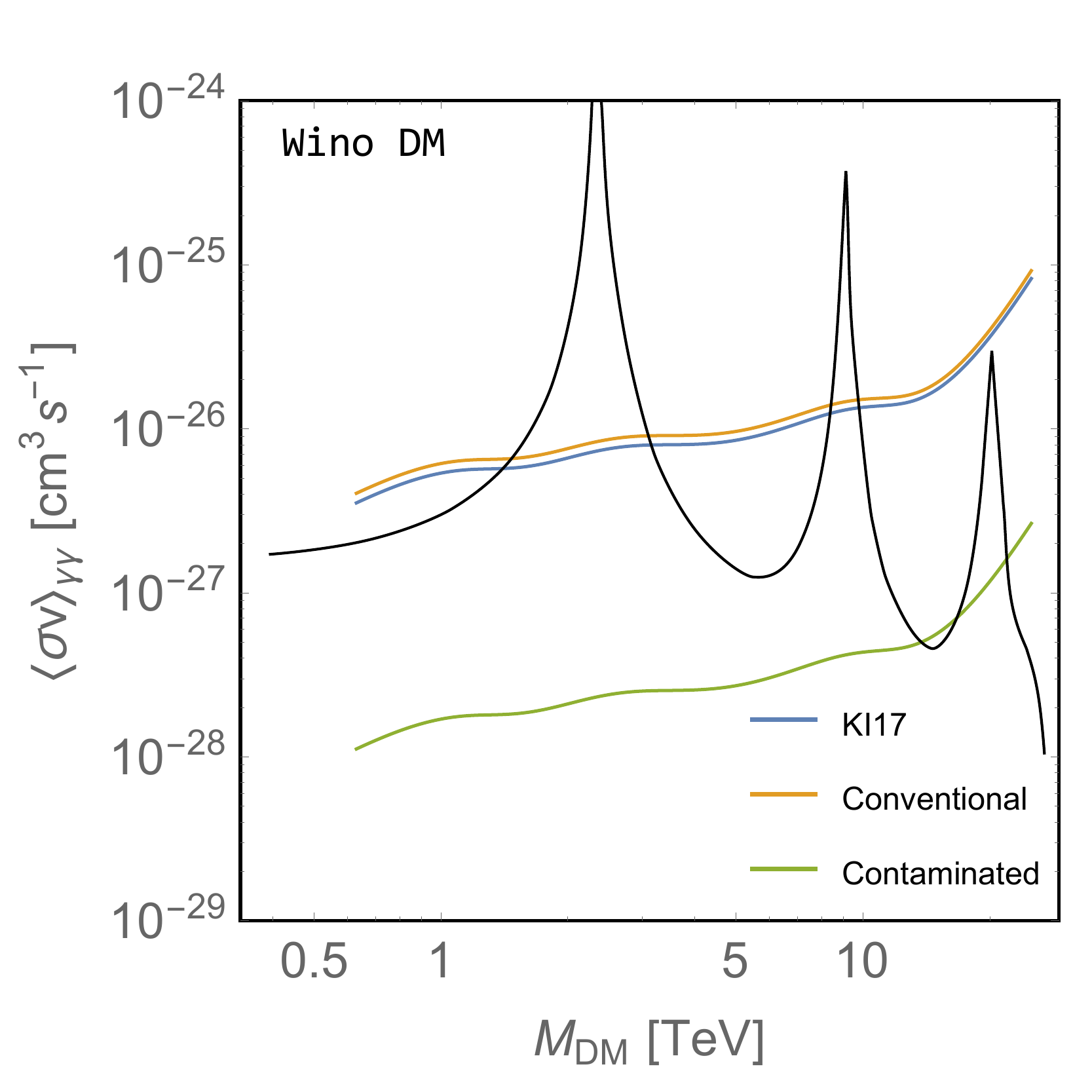}
\includegraphics[width=70mm,angle=0]{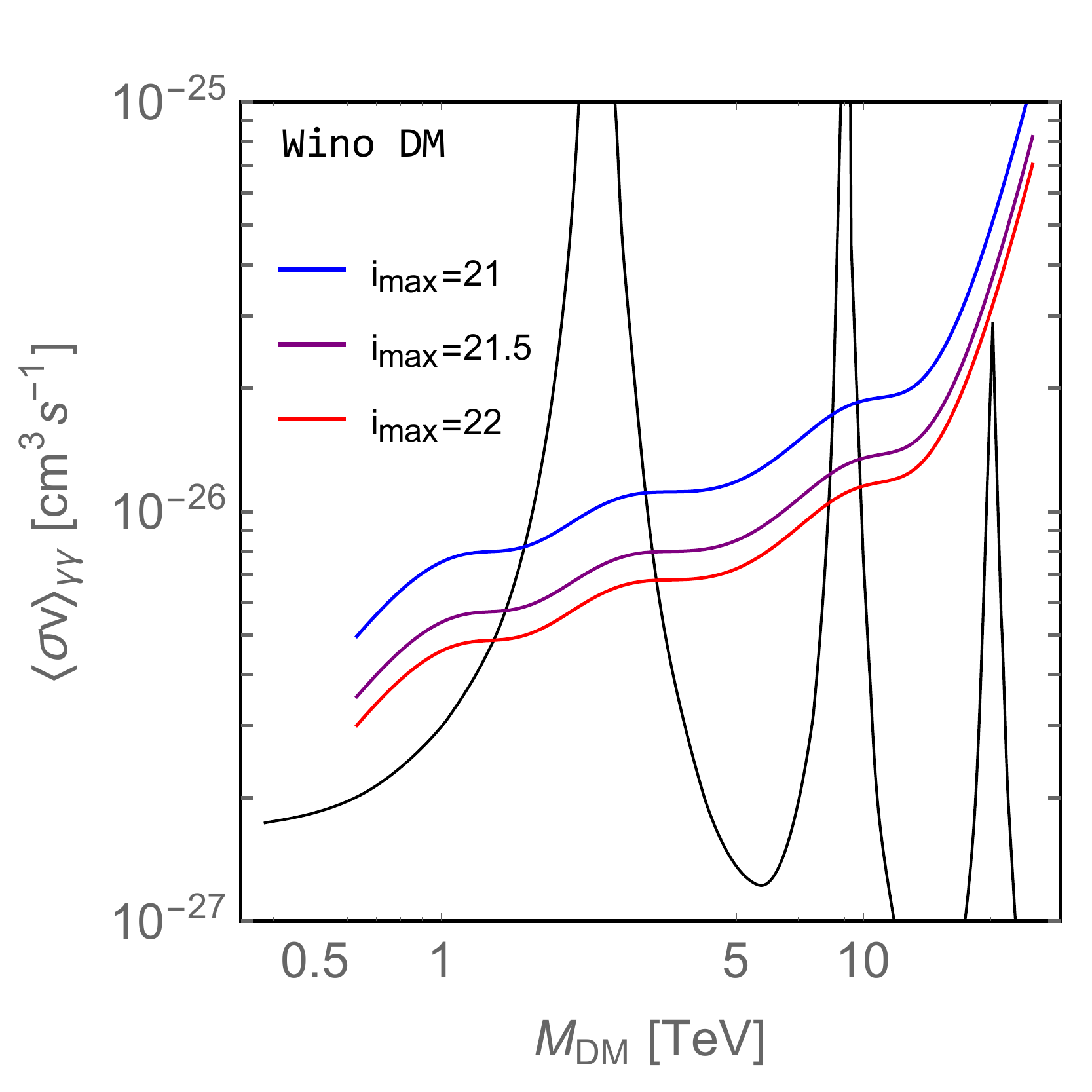}
\end{center}
\caption{\small\sl
Sensitivity lines of Wino DM annihilating into two photons $\langle\sigma v\rangle_{\gamma\gamma}$.
All coloured lines are achieved by combined likelihood analysis of 50 hours observation for each UFD.
The black line is the  photon cross section of Wino DM extracted by \citet{Lefranc2016}.
 {\bf Upper panel:}
 Sensitivity lines at $\imax = 21.5$.
 Each line assumes the $J$-factor values
 reproduced by the \KI, \MS and \Cont analysis (blue, orange and green, {respectively}). 
 {\bf Lower panel:}
 Sensitivity lines achieved by the \KI analysis at $\imax = 21,\,21.5$, and $22$ (blue, purple and red).
}
 \label{fig:snss}
\end{figure*}
%
In this appendix, 
we discuss the sensitivity of the future indirect dark matter detection observing gamma-rays from dSphs, 
in particular, focusing on how it depends on the median values and uncertainties of J-factors obtained in section 4.
In order to make our analysis concrete, we consider the Cherenkov Telescope Array project \citep{CTAO2017}
and the wino DM as a gamma-ray observatory and a DM candidate, respectively.

Wino DM is one of the most attractive WIMP candidates, 
where it is introduced in the supersymmetric extension of the standard model as the superpartner of the neural weak gauge boson. 
The wino DM attracts many attentions at present, 
because it is predicted in the anomaly-mediated supersymmetry breaking scenario explaining the Higgs boson mass of 125\,GeV 
as well as non-observation of new physics signals at collider experiments \citep{PhysRevD.85.095011}. 
Its mass is predicted to be \Order{1}\,TeV 
and its annihilation cross section is boosted by the so-called the Sommerfeld effect \citep{PhysRevLett.92.031303},
so that the indirect dark matter detection is expected to work very efficiently to detect the wino DM. 
We refer \citet{Lefranc2016} and \citet{Cirelli2011} for the branching fraction of each annihilation channel 
and the corresponding fragmentation function $(dN_\gamma /dE)_f$, respectively, to calculate the photon flux in Eq.\,(\ref{eq:fluxformula}).


CTA will provide significantly improved sensitivity to WIMP DM with its high angular resolution and its wide energy coverage.
Here we briefly review the CTA analysis in \citet{Lefranc2016}. The number of photon count in $i$-th energy bin $N^i_\gamma$ is given by $N^i_\gamma=N^i_\text{sg}+N^i_\text{bg}$.
Here $N^i_\text{sg}$ denotes the number of the signal photon in the $i$-th energy bin $\Delta E_i$, given by 
\begin{eqnarray}
N^i_\text{sg}  = T_\text{obs}\times\int_{\Delta E_i}\!\!{dE'}\int\!\!{dE_\gamma}\ 
	\mathcal{A}_\text{eff}(E_\gamma)\,\Phi(E_\gamma,\Delta\Omega)\,\mathcal{R}(E_\gamma,E')\ .
\end{eqnarray}
{Here, $T_\text{obs}$ is the observation time for each UFD, which is assumed to be $T_\text{obs}$ = 50 hours as a benchmark.}
{The differential flux $\Phi$ is the one obtained by Eq.\,(\ref{eq:fluxformula}),}
{The effective area $\mathcal{A}_\text{eff}(E_\gamma)$
and the energy resolution of $\mathcal{R}(E_\gamma,E')\equiv\mathcal{G}[E';E_\gamma,\sqrt{8\ln(2)}\delta_\text{res}(E_\gamma)]$ are given in}
\citet{CTAO2017}.
The background rate, $\nu_i$, is also given in \citet{CTAO2017}, from which we obtain the number of the background photon $N_\text{bg}^i$ by multiplying $T_\text{obs}$.
{In our analysis, the size of each energy bin is the same as that of the background rate in the reference.}

The likelihood function of the indirect detection for a specific dSph (indexed with $j$) is given by
\begin{eqnarray}
	\mathcal{L}_j(\langle \sigma v \rangle)&=& \left.\max_{J} \left(\prod_i \frac{{N^i_\gamma}^{N^i_\text{obs}}}{N^i_\text{obs}}e^{-N^i_\gamma}\right)
		\,\times\, \pi(J)\right|_{J=J_j},\\
	\pi(J)&=&\frac{\mathcal{G}(\log_{10} J; \log_{10} J_\text{mean}, \delta\log_{10}J)}{\ln(10)J}\ ,
	\label{eq:CTAlikelihood}
\end{eqnarray}
where $N^i_\text{obs}$ denotes the photon number observed in $i$-th energy bin.
{To estimate the mean sensitivity of the annihilation cross section for the null observation, we assume $N^i_\text{obs}={N^i_\text{bg}}$.}
{The factor $\pi(J)$ represents the uncertainty of the $J$-factor which is discussed in this paper.}
Total likelihood function is given by the product of each likelihood function $\mathcal{L}(\langle \sigma v \rangle) = \prod_{j}\mathcal{L}_j(\langle \sigma v \rangle)$.
Then we perform the statistical test with the condition 
$\chi^2\equiv-2\ln[\mathcal{L}(\langle \sigma v \rangle)/\mathcal{L}_\text{max}]<2.71$ (95\% confidence level).
{The sensitivity line, i.e. the prospected upper limit on $\langle \sigma v \rangle$, is obtained by
 $\chi^2 = 2.71$ as the function of $M_\text{DM}$.}



%
The upper panel of Fig.\,\ref{fig:snss} shows 
{the} sensitivity lines of photon cross section $\langle\sigma v\rangle_{\gamma\gamma}$ {for the case of $\imax=21.5$}.
{For the illustration, we convert the sensitivity lines $\langle\sigma v\rangle$, obtained by $\chi^2=2.71$, into 
$\langle\sigma v\rangle_{\gamma\gamma}= b_{\gamma\gamma}\langle\sigma v\rangle$.}%
\footnote{
For the ''pure" photon model where $b_{\gamma\gamma}^\text{pure}=1$, for instance,
its sensitivity $\langle\sigma v\rangle_{\gamma\gamma}^\text{pure}$
is obtained by
$(1+b_{\gamma Z}^\text{Wino}/2b_{\gamma\gamma}^\text{Wino})
\langle\sigma v\rangle_{\gamma\gamma}^\text{Wino}$,
where $\langle\sigma v\rangle_{\gamma\gamma}^\text{Wino}$ is the sensitivity of Wino DM in Fig.\,\ref{fig:snss}.
This reinterpretation is verified
because the continuum spectrum of Wino DM barely affect to its sensitivity lines.
}
{The sensitivity line obtained by the \Cont approach is about 100 times severer than {the} other methods.}
{The sensitivity lines simply {reflect} the estimated value of the $J$-factor,}
{which} {show} the importance of a careful estimation of the $J$-factors, 
since otherwise the dark matter model will be constrained too aggressively. 
{The \MS and the \KI approaches avoid such a problem and, when all of the four dSphs are taken into account in the likelihood, there is no significant difference between them at the level of the present observational depth.} 
{We here note that in case of the observation of \Seg only, even which is the most promising target due to its large $J$-factor, the difference between the \MS and the \KI approaches is more significant.}
The difference will {also} appear when the {$J$}-factors are estimated at deeper observation, as can be expected from the {$J$}-factors in Fig.\,\ref{fig:Jfig}.
In the lower panel of Fig.\,\ref{fig:snss} we show the improvement of the sensitivity with the \KI method by increasing the observation {depth}.
From the view point of the thermal Wino dark matter search where its mass is predicted to be about 3\,TeV\, \citep{Hisano:2006nn}, 
it will be crucial to choose the observational depth at around $\imax = 21.5$, as can be seen in the panel.

\red{
\section{Effect of the truncation radius}
\begin{figure*}
\centering
\includegraphics[width=55mm]{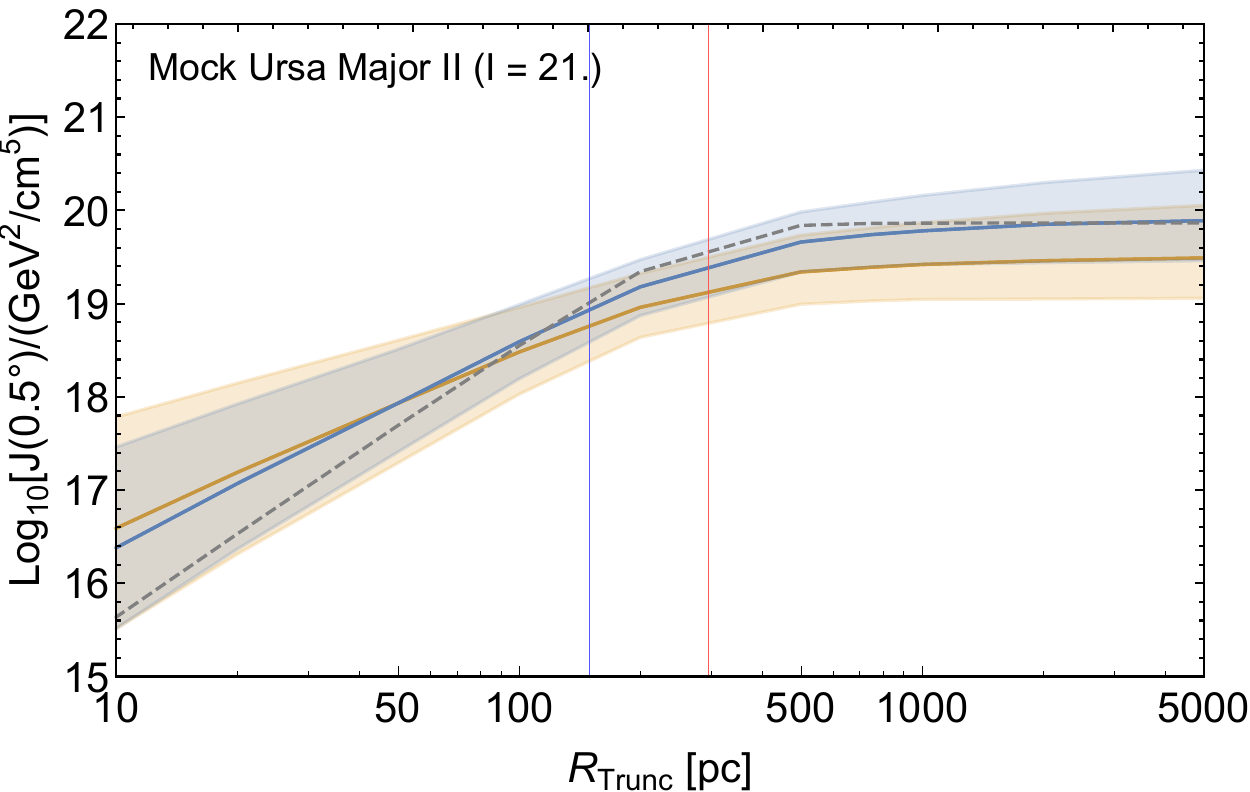}
\includegraphics[width=55mm]{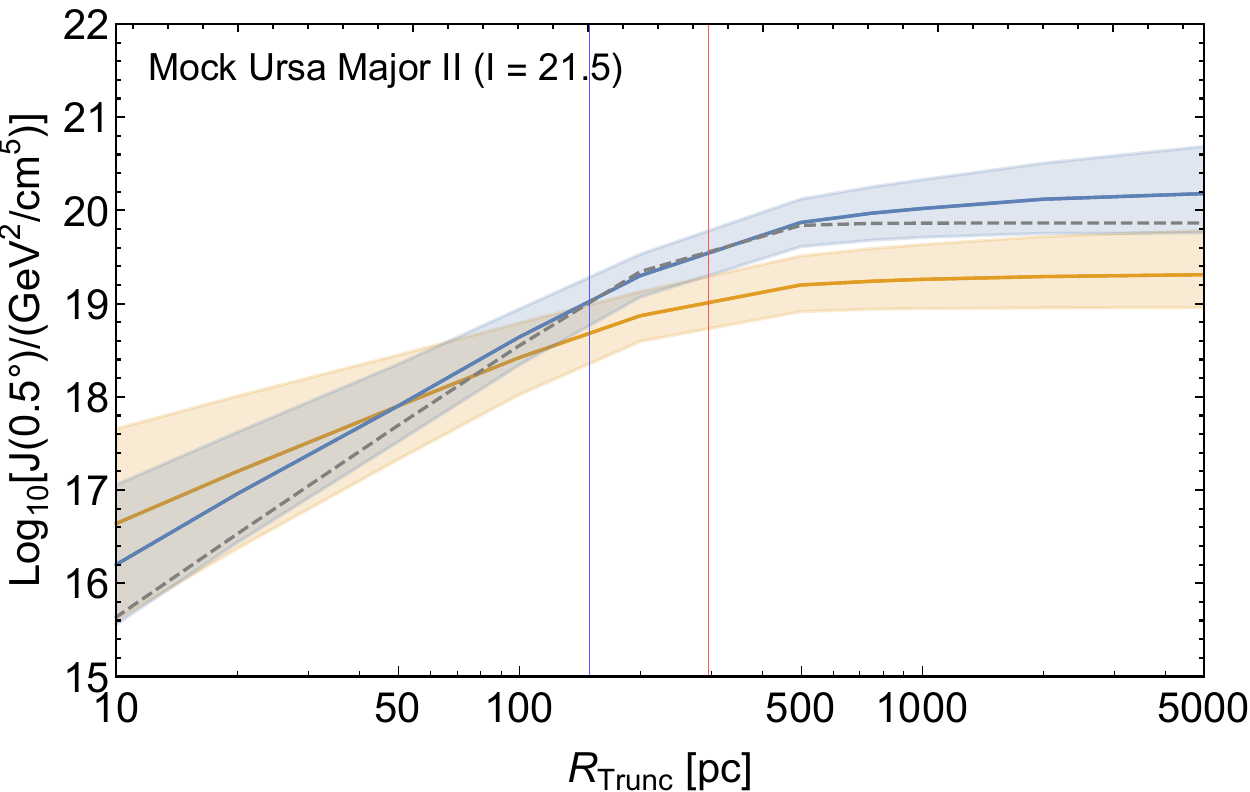}
\includegraphics[width=55mm]{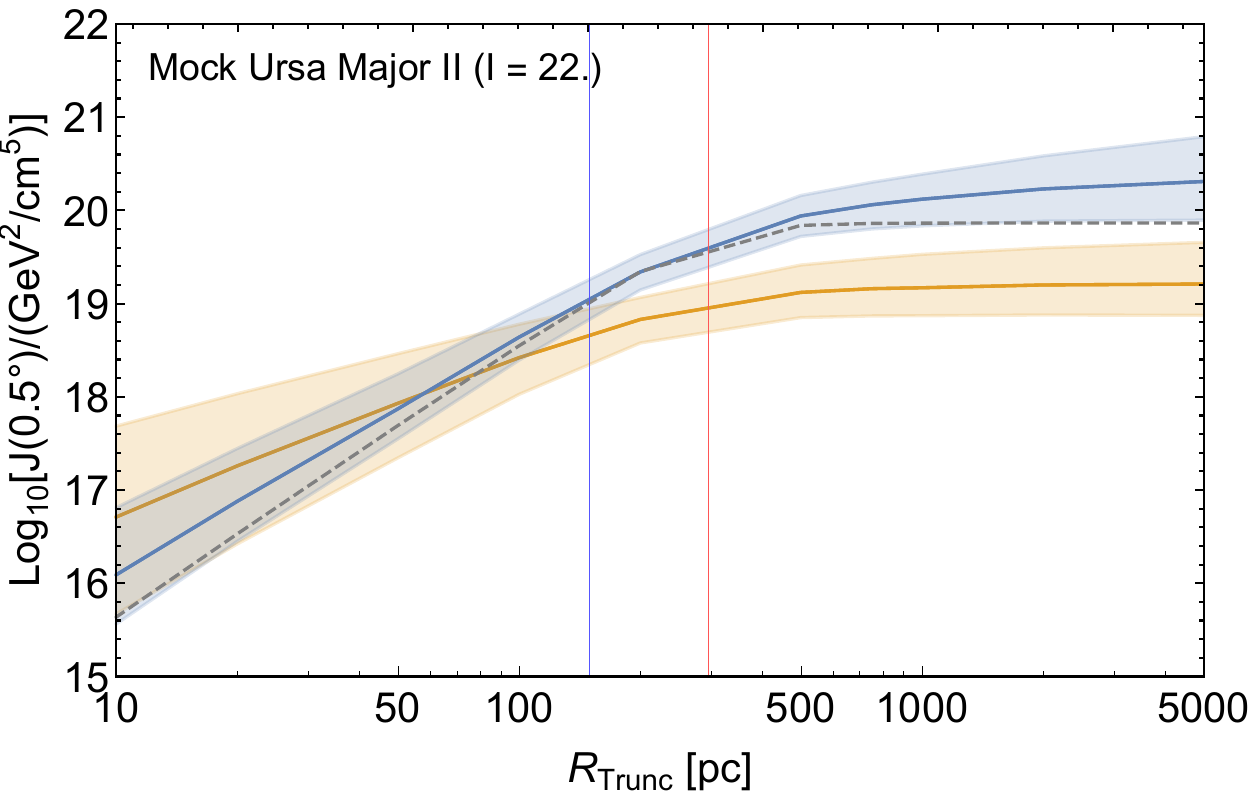}\\
\includegraphics[width=55mm]{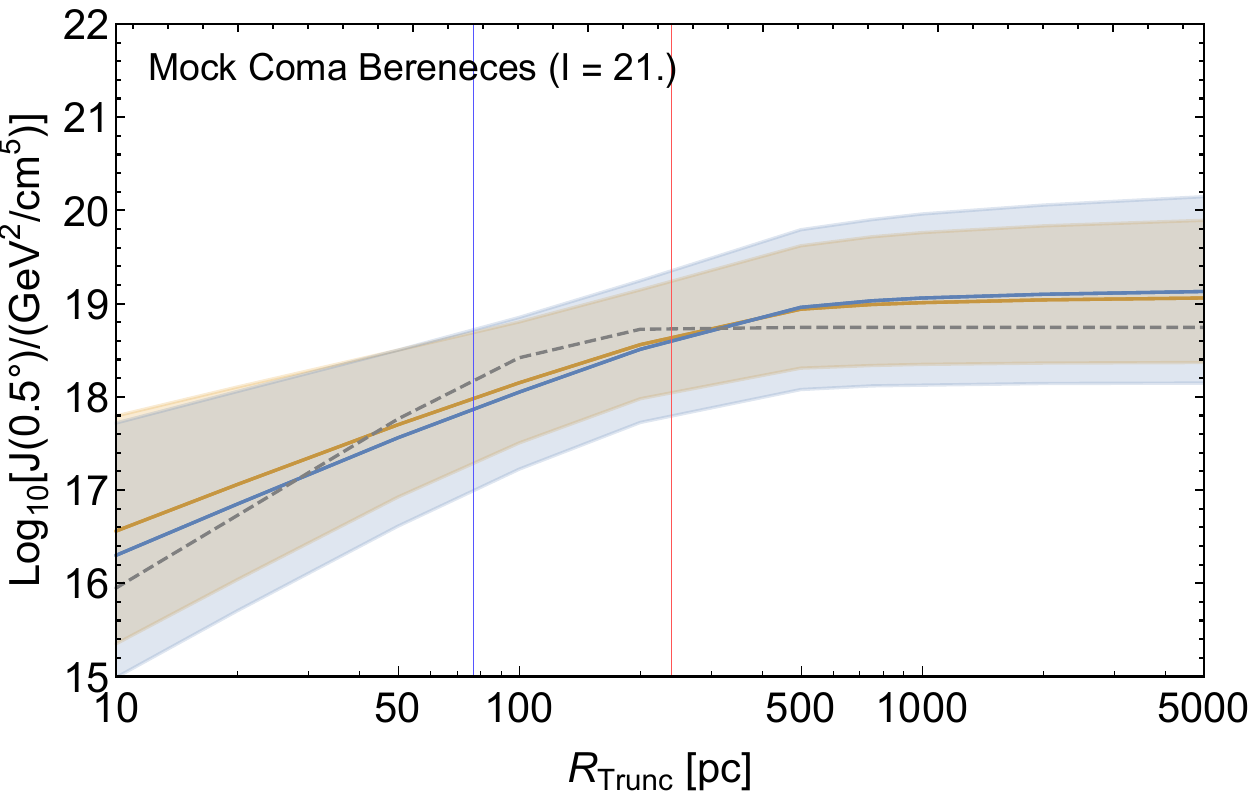}
\includegraphics[width=55mm]{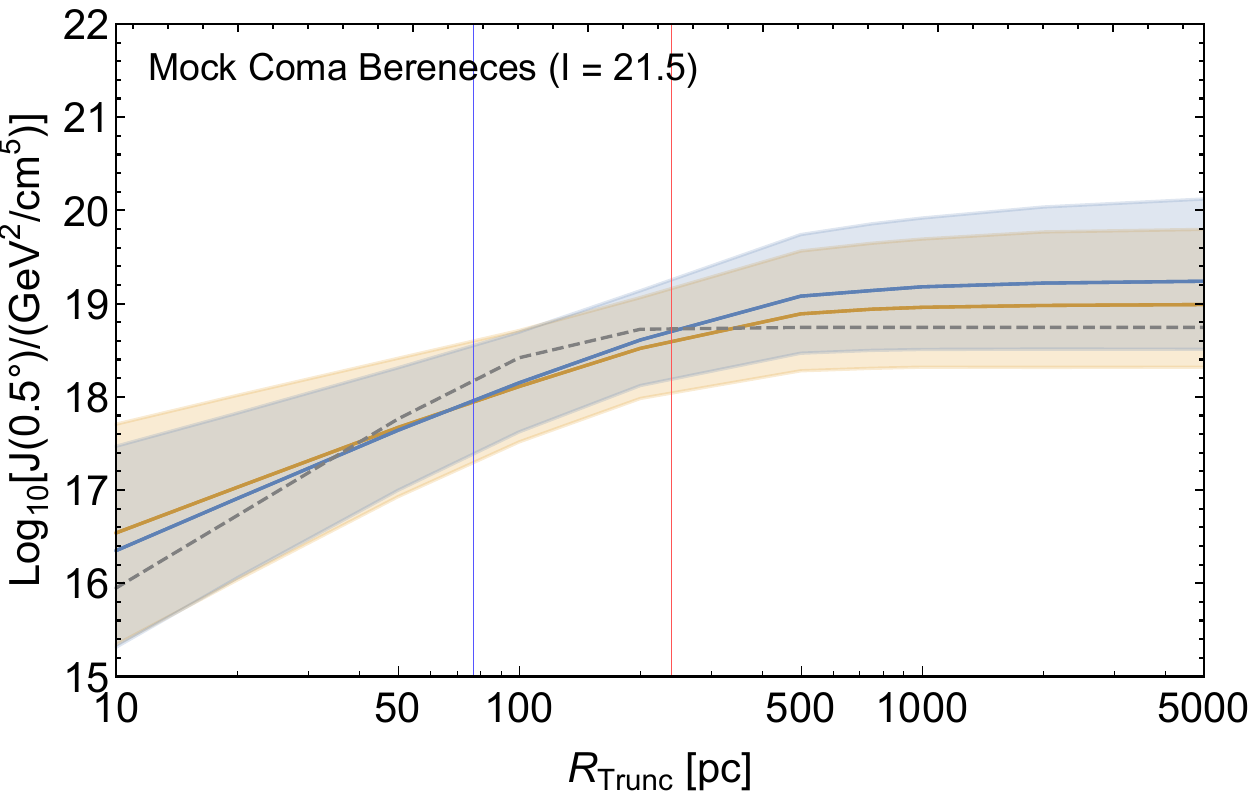}
\includegraphics[width=55mm]{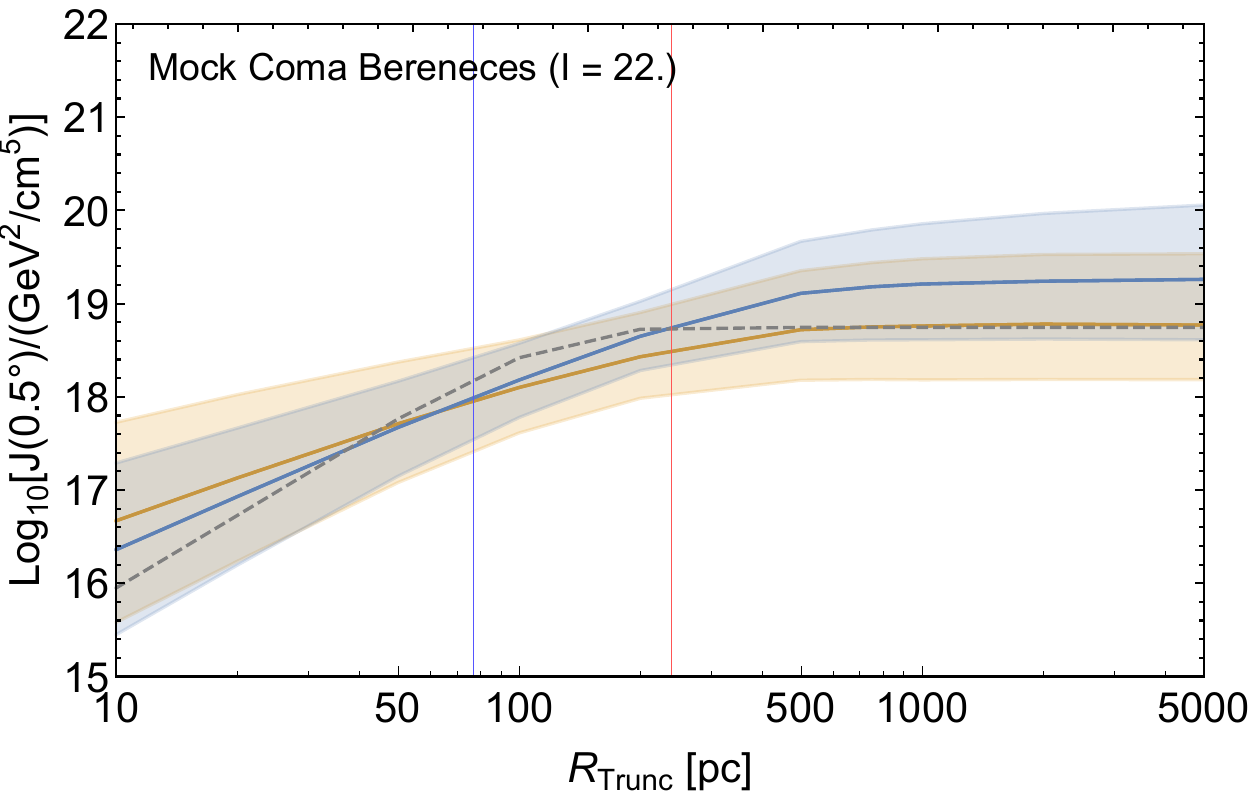}\\
\includegraphics[width=55mm]{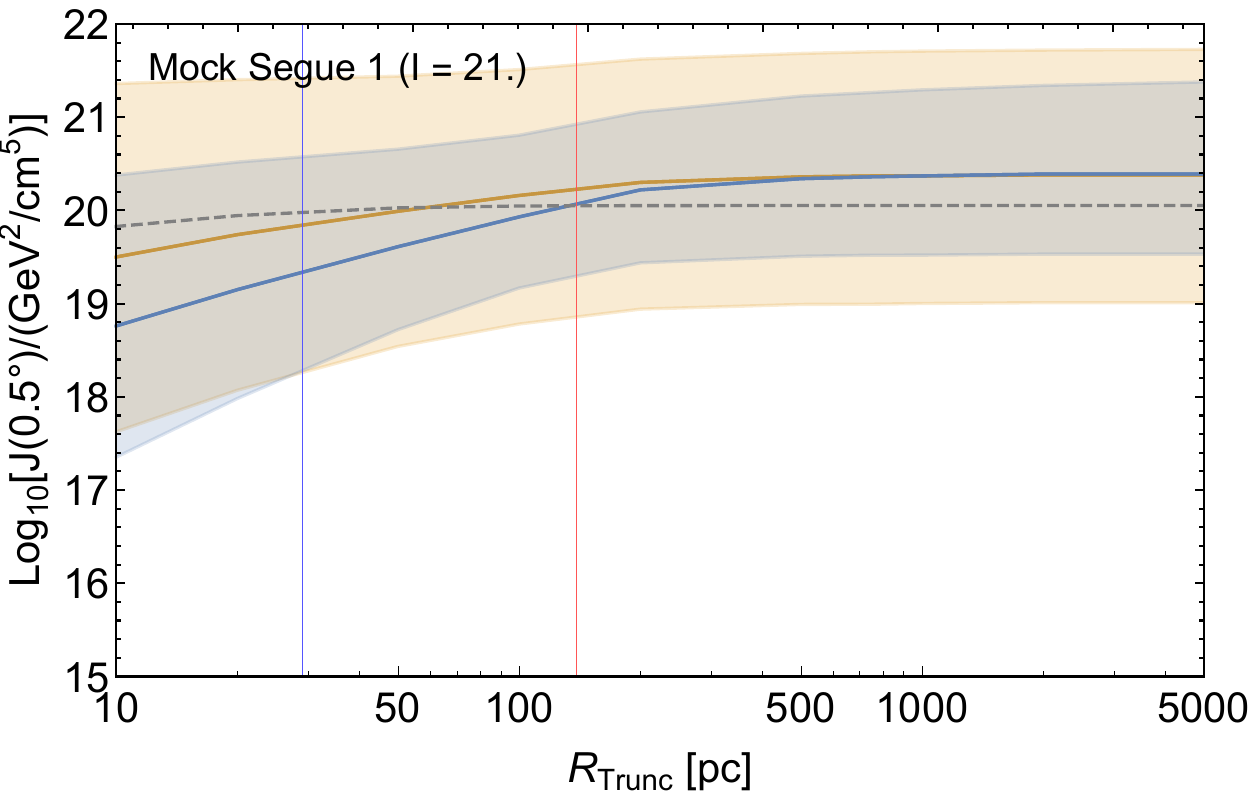}
\includegraphics[width=55mm]{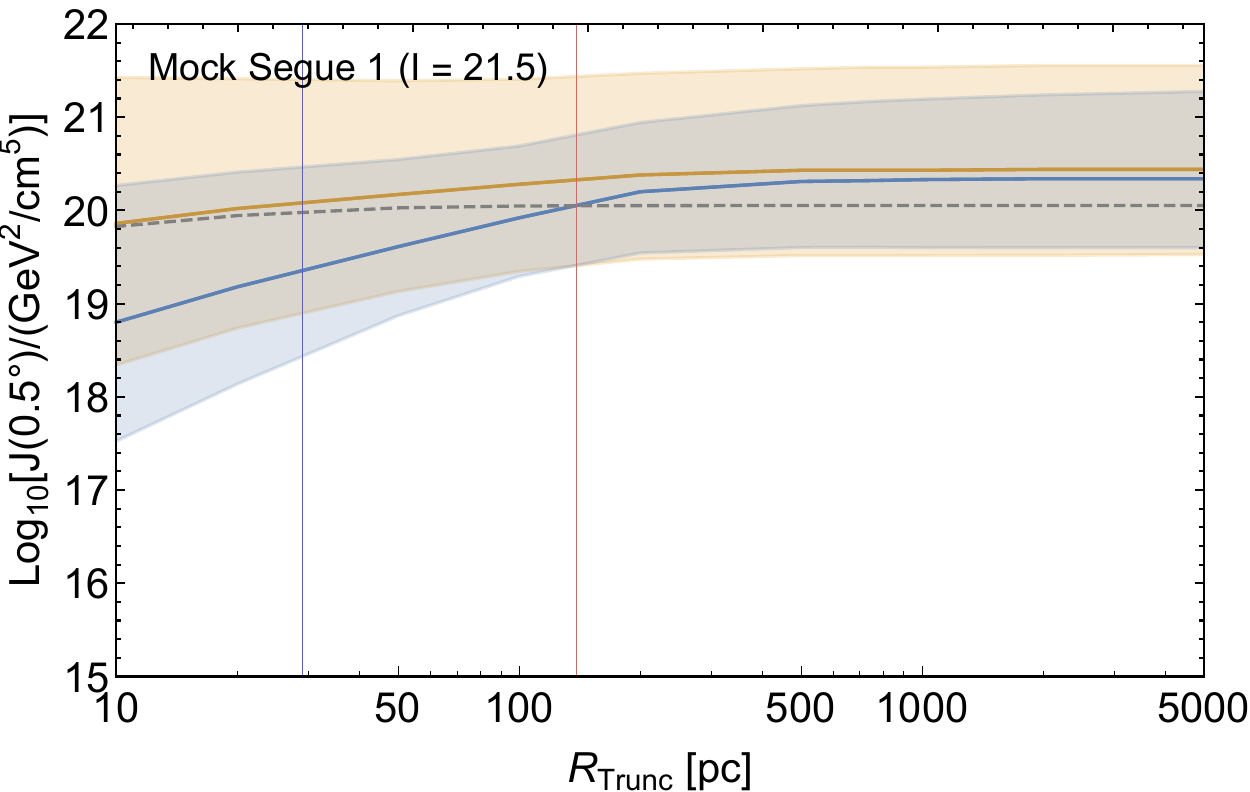}
\includegraphics[width=55mm]{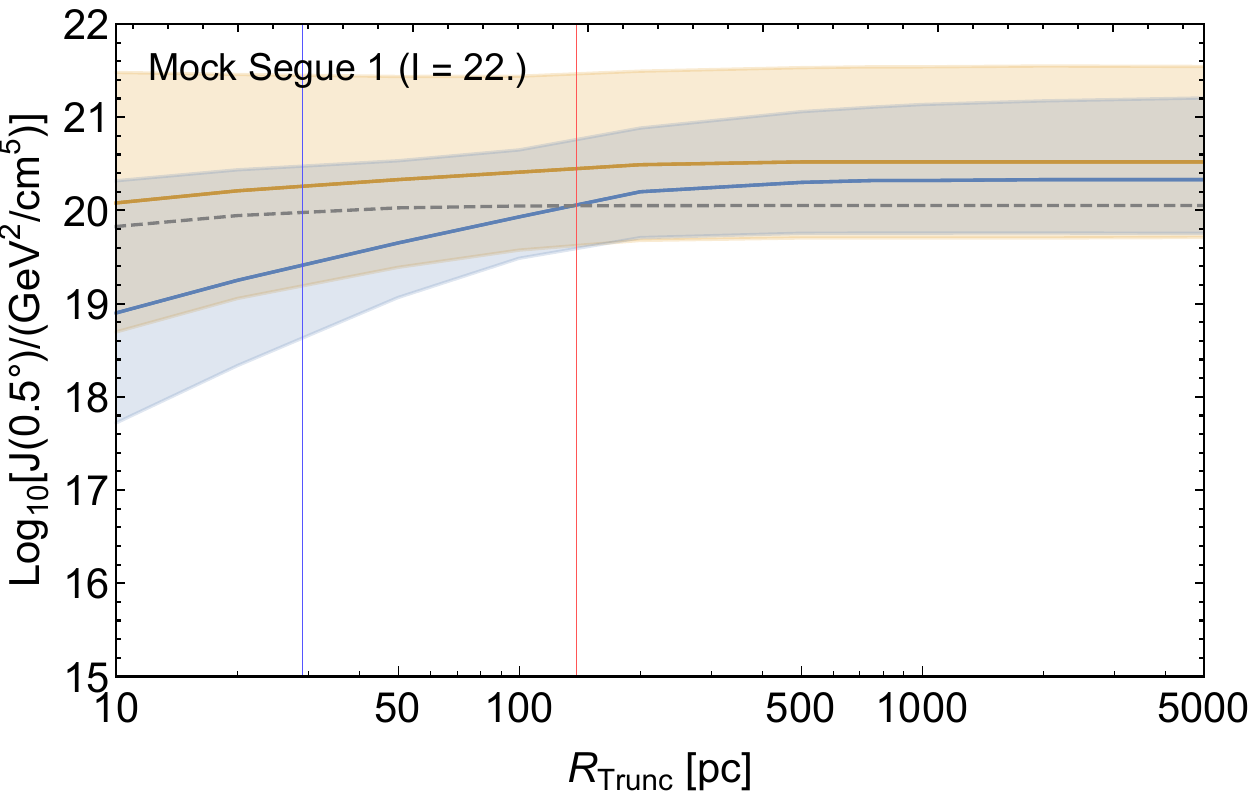}\\
\includegraphics[width=55mm]{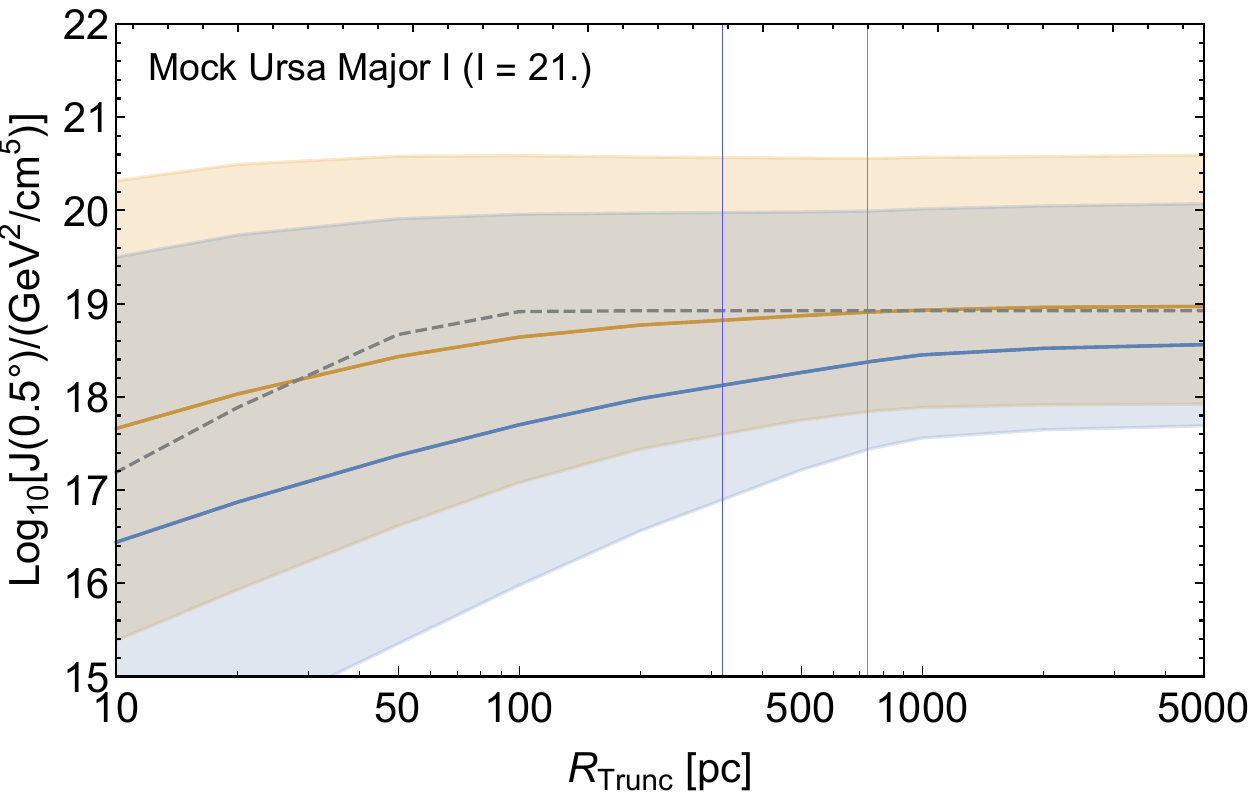}
\includegraphics[width=55mm]{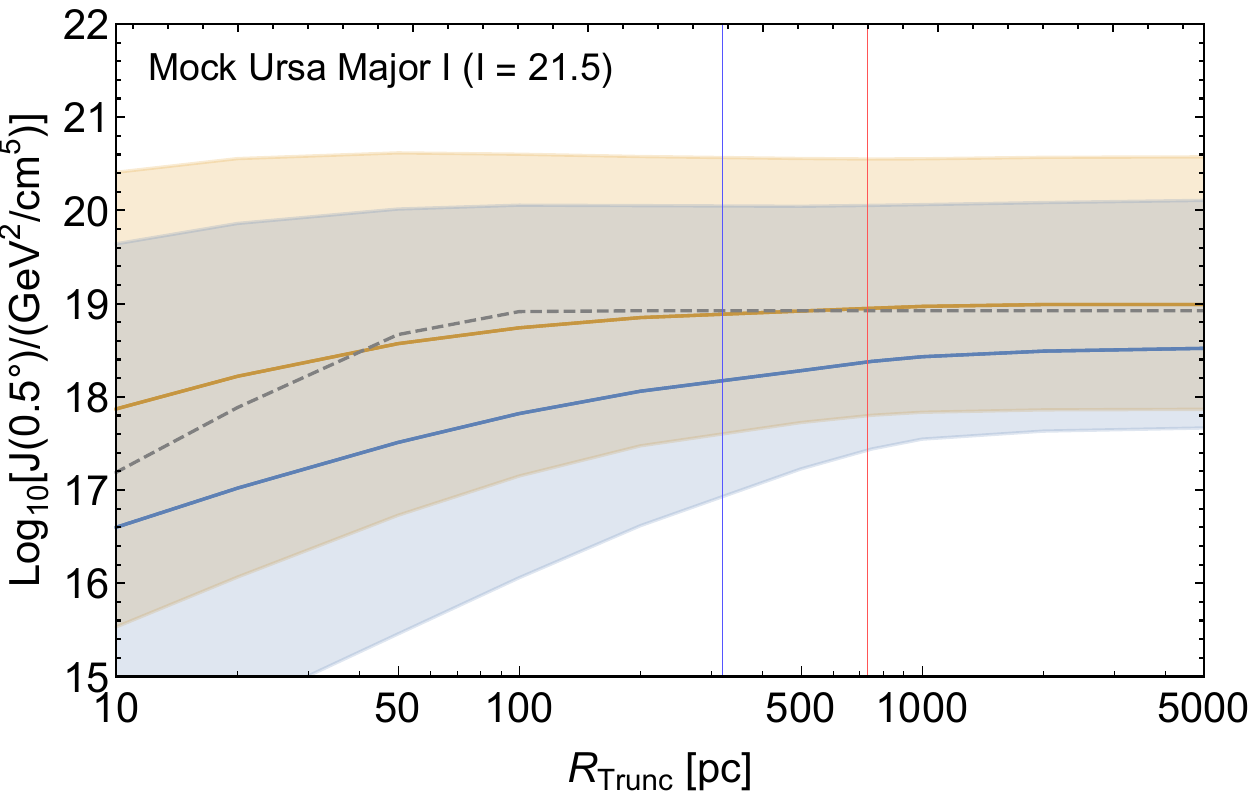}
\includegraphics[width=55mm]{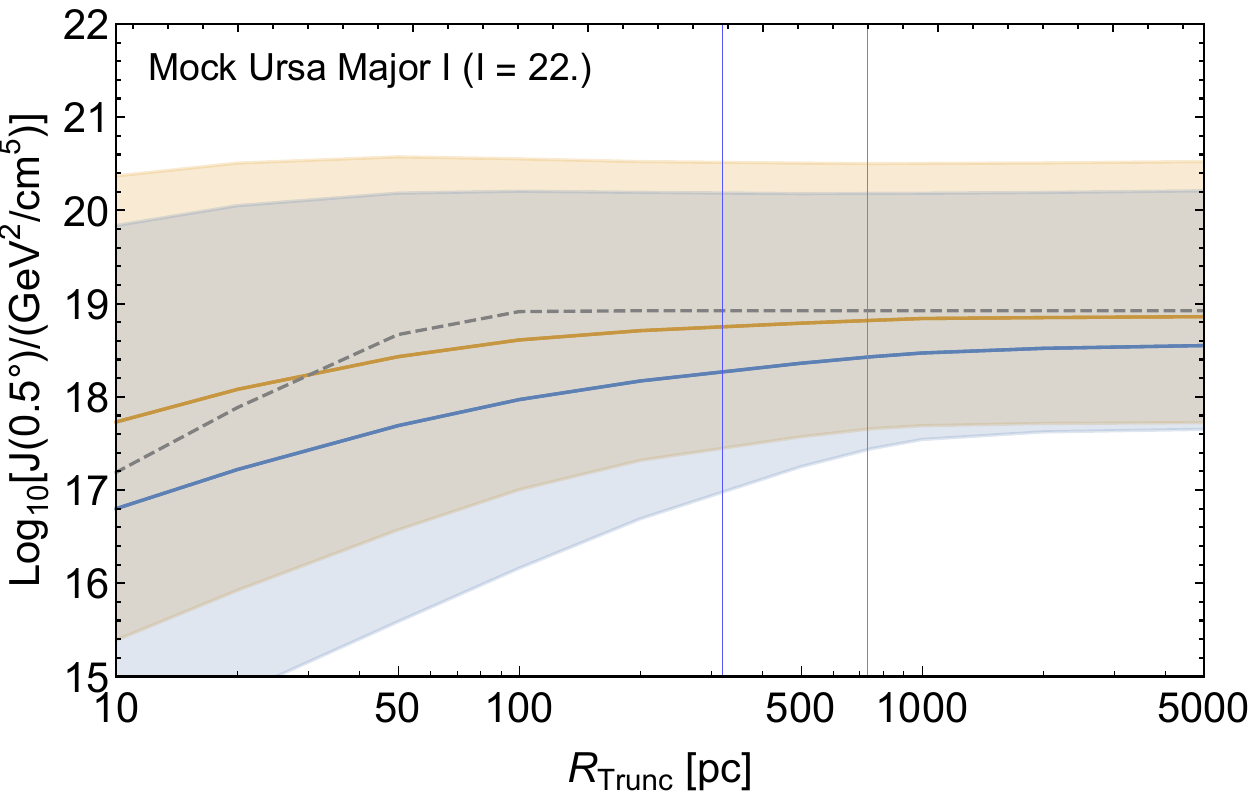}
\caption{\red{The truncation radius ($R_\text{trunc}$) dependence of the $J$-factor estimation for each $i_\text{max} = 21,21.5$, or $22$ (left/centre/right) at the four dSphs.
The input values are shown by the dotted curves.
The blue (orange) solid curves and shaded area show the median value of the $J$-factor and its 68\% quantile for \KI (\MS). The vertical blue (red) lines correspond to $r_e$ ($r_\text{max}$ in Table \ref{tb:Cut}) of each dSph.
}}
\label{fig:trunc}
\end{figure*}
%
The truncation radius dependence of the $J$-factor is shown in Fig. \ref{fig:trunc} for each $i_\text{max} = 21,21.5$, or $22$ at the four dSphs.
The figure shows that the both \MS and \KI analyses reproduce the input $J$-factor well for any $R_\text{trunc}$.
\footnote{The only exception is seen for the case of \UII at the \MS analysis (top right panel).}
Moreover, it can also be seen that the value of $J$-factor is not alter significantly when $r_\text{max}$ is similar to or larger than $r_\text{max}$.
We hence set the truncation radius of the dark matter halo to $r_\text{max}$ in Table \ref{tb:Cut} in our analysis of this paper.
}

\bsp	
\label{lastpage}
\end{document}